\documentclass[apj]{emulateapj}
\usepackage{apjfonts}

\gdef\kms{km\,s$^{-1}$}
\gdef\msun{$M_{\odot}$}
\lefthead{van Dokkum}
\righthead{Evolution of the IMF}
\slugcomment{Astrophysical Journal, in press}
\begin{document}

\title{Evidence of Cosmic Evolution of the Stellar Initial Mass Function}
\altaffilmark{1}

\author{Pieter~G.~van Dokkum}
\affil{Department of Astronomy, Yale University,
New Haven, CT 06520-8101}

\altaffiltext{1}
{Based on observations with the NASA/ESA {\em Hubble Space
Telescope}, obtained at the Space Telescope Science Institute, which
is operated by AURA, Inc., under NASA contract NAS 5--26555.}

\begin{abstract}

Theoretical arguments and indirect observational evidence suggest that the
stellar initial mass function (IMF) may evolve with time, such that
it is more weighted toward high mass stars at higher redshift.
Here we test this idea by comparing the rate of luminosity evolution
of massive early-type galaxies in clusters at $0.02\leq z\leq 0.83$
to the rate of their color evolution.  A combined fit
to the rest-frame $U-V$ color evolution
and the previously measured evolution of the $M/L_B$ ratio
gives $x = -0.3^{+0.4}_{-0.7}$ for the logarithmic slope
of the IMF in the region around 1\,\msun, significantly
flatter than the present-day value in the Milky Way disk
of $x= 1.3\pm 0.3$. The best-fitting luminosity-weighted formation
redshift of the stars in massive cluster galaxies is $3.7^{+2.3}_{-0.8}$,
and a possible
interpretation is that the characteristic mass $m_c$ had a value
of $\sim 2$\,\msun\ at $z\sim 4$ (compared to $m_c \sim 0.1$\,\msun\
today), in qualitative agreement with models in which the characteristic
mass is a function of the Jeans mass in molecular clouds.
Such a ``bottom-light''
IMF for massive cluster galaxies
has significant implications for the interpretation
of measurements of galaxy formation and evolution.
Applying a simple form of IMF evolution 
to literature data, we find that
the volume-averaged star formation rate at high redshift
may have been overestimated
(by a factor of $3-4$ at $z> 4$), and
the cosmic star formation history may have
a fairly well-defined peak at $z\sim 1.5$.
The $M/L_V$ ratios of galaxies are less affected than
their star formation rates, and future data on the stellar mass
density at $z>3$ will provide further constraints on IMF evolution.
The formal errors likely underestimate the uncertainties,
and confirmation of these results requires a larger sample of
clusters and the inclusion of redder rest-frame colors in the
analysis.

\end{abstract}


\section{Introduction}

The form of the stellar initial mass function (IMF) is of
fundamental importance for many areas of astrophysics and
a topic of considerable debate
(see, e.g., {Schmidt} 1959; {Miller} \& {Scalo} 1979; {Scalo} 1986; {Larson} 1998, 2003; {Kroupa} 2002; {Chabrier} 2003, for reviews).
Measurements of the IMF are difficult and somewhat model-dependent
as they require the conversion of the observed present-day luminosity
function of a stellar population to its mass function at birth.
Best estimates for the Galactic disk suggest that the IMF has a
powerlaw slope at $m\gtrsim 1$\,\msun, and  turns
over at lower masses ({Kroupa} 2001; {Chabrier} 2003). This turnover
can be modeled by a broken powerlaw
({Kroupa} 2001) or by a log-normal distribution with a characteristic
mass $m_c$ ({Chabrier} 2003). The value of $m_c$ is $\sim 0.1$\,\msun\
in the disk of the Milky Way, with considerable uncertainty.
The powerlaw slope at high masses
is probably close to the {Salpeter} (1955) value of $x=1.35$, with an
uncertainty of $\sim 0.3$ ({Scalo} 1986; {Chabrier} 2003).

Although there is no  direct
evidence for dramatic variations of the IMF within the
present-day Milky Way
disk (e.g., {Kroupa} 2001; {Chabrier} 2003),
this does not preclude variations with time, metallicity, and/or
environment.  In particular,
{Larson} (1998, 2005) has argued that the characteristic turnover
mass may be largely determined by the thermal Jeans mass,
which strongly depends on temperature ($\propto T^{3/2}$ at fixed
density).
In the context of this model
one might expect that heating by ambient far-infrared radiation
would disfavor
the formation of low mass stars in extreme
environments, such as in super star clusters
and in the center of the Milky Way.
Other models emphasize the role of turbulence as opposed
to temperature in determining
the distribution of protostellar clumps (e.g., {Padoan} \& {Nordlund} 2002),
and in these models the role of the environment may be less direct
(see {McKee} \& {Ostriker} [2007]
for a recent review of various models to explain the
characteristics of the IMF).

Observations may support the notion of a top-heavy (or ``bottom-light'')
IMF in extreme
environments.
Some young super star clusters in M82 appear to have a
top-heavy mass function (e.g., {Rieke} {et~al.} 1993; {McCrady},
{Gilbert}, \& {Graham} 2003), as do clusters in the Galactic center
region (e.g., {Figer} {et~al.} 1999; {Stolte} {et~al.} 2005; {Maness}
{et~al.} 2007).
The interpretation of observed mass functions
is complicated by
dynamical effects, which tend to make the mass function more top-heavy
over time, in particular in the central regions of star clusters
(see, e.g., {McCrady} {et~al.} 2003; {McCrady}, {Graham}, \&
{Vacca} 2005; {Kim} {et~al.} 2006).
Recently Harayama, Eisenhauer, \& Martins (2007)
studied the IMF of NGC 3603, one of the most massive Galactic
star-forming regions, out to large radii
and conclude that its IMF is substantially
flatter than Salpeter for masses $0.4 -20$\,\msun.

The IMF may also depend on redshift.
At earlier times star formation presumably occurred more often
in a burst mode than in a relatively gradual ``disk'' mode
(e.g., {Steidel} {et~al.} 1996; {Blain} {et~al.} 1999b; {Lacey} {et~al.} 2007), which means that the IMF could generally be more
skewed toward high mass stars at redshifts 1--3 and beyond.
Furthermore, the average metallicity in star forming clouds
was lower at higher redshift, which may have led to an
extremely top-heavy IMF for the first generation of stars
(e.g., {Abel}, {Bryan}, \& {Norman} 2002; {Bromm}, {Coppi}, \& {Larson} 2002). Finally,
the cosmic microwave background (CMB) radiation sets a floor to
the ambient temperature, and hence the Jeans mass, which
scales with $(1+z)$. Beyond $z\sim 2$ the CMB temperature
exceeds the typical temperatures of dense prestellar cores
in Galactic
molecular clouds (e.g. {Evans} {et~al.} 2001; {Tafalla} {et~al.} 2004).
Therefore, at sufficiently high
redshift the characteristic mass may be expected to evolve
roughly as $m_c \propto (1+z)^{3/2}$, leading to IMFs which
have a reduced fraction of low mass stars ({Larson} 1998). 
The effects of the CMB are even more pronounced when
its influence on the pressure in star-forming clouds is taken
into account, and {Larson} (2005) suggests that
at $z=5$ the characteristic mass may be higher than
today's value by as much as an order of magnitude.
Such rapid evolution of the IMF
would have important consequences
for determinations of masses and star formation rates of distant
galaxies, and for measurements of evolution in these properties.

It is very difficult to constrain the IMF at early times
directly, as the light of high redshift galaxies is completely
dominated by massive stars.
The extremely blue rest-frame UV colors of
galaxies at $z\sim 6$ may imply a top-heavy IMF
({Stanway}, {McMahon}, \&  {Bunker} 2005), although this is just one
of several possible explanations.
From observations of a lensed Lyman break galaxy
there is some evidence that the {\em slope} of the IMF at $z\sim 3$
is similar
to the Salpeter value at the high mass end ({Pettini} {et~al.} 2000), but
there is essentially no information on stars with masses near or
below 1\,\msun. 

Fortunately, the form of the high redshift IMF has implications for the
properties of galaxies at much lower redshift,
as all stars with masses $\lesssim 0.8$\,\msun\ that formed
in the history of the Universe are still with us today.
Tumlinson (2007) finds that the properties of carbon-enhanced
metal-poor stars in our Galaxy are best explained with a
relatively high number of stars in the mass range
1--8\,\msun\  at high redshift.
Various other constraints obtained from galaxies at low redshift
(including our own) are reviewed in {Chabrier} (2003).
Of particular interest are
the stellar populations of massive early-type galaxies,
as they are very homogeneous and should reflect conditions in
star forming regions at $z> 2$. 
Recently, {Cappellari} {et~al.} (2006) used the
kinematics of elliptical galaxies to constrain
the IMF, as the dynamical $M/L$ ratio provides an upper limit to the
amount of mass that can be locked up in low mass stars.
Current data
appear to rule out a {Salpeter} (1955) (or steeper) IMF, but are
consistent with {Kroupa} (2001) and {Chabrier} (2003) IMFs
({Cappellari} {et~al.} 2006).

In this paper, we provide new constraints on the IMF at
high redshift by comparing the {\em evolution} of
the $M/L$ ratios of early-type galaxies
to their color evolution. This method was first suggested by
{Tinsley} (1980), but data of sufficient accuracy are only now becoming
available. The method is sensitive to the IMF in the important
mass range around 1\,\msun, where the effects of an evolving
characteristic mass might be expected to manifest themselves.

A plan of the paper follows. In \S\,2 a relation between
color evolution, luminosity evolution, and the logarithmic
slope of the IMF $x$ is derived using stellar population synthesis
models. In \S\,3 published data and archival HST images of
galaxy clusters are used to construct the redshift evolution of the $U-V$
color-mass relation. In \S\,4 the color evolution from
\S\,3 is combined with the previously measured evolution
of the $M/L_B$ ratio. The relations from \S\,2 are then
used to derive constraints on the IMF slope $x$ from
the combined color and luminosity evolution. Section 5
is devoted to the (many) systematic uncertainties in the methodology and
in the data, and \S\,6 asks whether our results are consistent
with other constraints on the stellar populations of massive
early-type galaxies. Although the constraints we derive in this paper
are subject to many uncertainties, it is interesting to explore
their consequences. In \S\,7 the fitting results of
\S\,4 are interpreted
in the context of an evolving characteristic mass $m_c$. The data on
cluster galaxies are combined with previous constraints on $m_c$
for globular clusters and submm-galaxies, and a simple form of
IMF evolution is proposed. This evolution is then
applied to literature data on the evolution of the volume-averaged
star formation rate and stellar mass density. The key
results are summarized in \S\,8.
We assume $\Omega_m=0.3$,
$\Omega_{\Lambda}=0.7$, and $H_0=71$\,\kms\,Mpc$^{-1}$ where needed.

\section{Methodology}
\label{method.sec}

\subsection{Effects of the IMF on Luminosity and Color Evolution}
\label{yy.sec}

The form of the IMF has a strong effect on the
evolution of the $M/L$ ratio of galaxies.
The luminosity of a stellar population is expected to evolve
even in the absence of star formation or mergers, due to the
fact that stars turn off the main sequence and die. The rate
of luminosity evolution in a given passband
is determined by a combination of
the lifetime of stars near the turn-off and the number of
these stars. The lifetime of turn-off stars is directly
determined by the age of the stellar population (with the
lifetime approximately equal to the age), and the number
of turn-off stars is directly determined by the IMF.
A flat, top-heavy IMF implies a relatively large number of short-lived
massive stars and rapid luminosity
evolution, whereas a steep IMF implies a relatively large
number of long-lived low mass stars and slow evolution.

The color evolution behaves differently, as it
is insensitive to the number of stars that evolves
off the main sequence over time. Instead it is
mainly driven by the location of the turn-off in the
Herzsprung-Russell (HR) diagram,
which does not depend on the IMF
but is determined by the age of the stellar population.
The IMF does influence the color evolution to some extent,
as a top-heavy IMF reduces the number of turn-off stars
with respect to the more luminous red giants. The location
of the giant branch evolves much less than the location of
the turn-off, and a flat IMF therefore
implies weaker color evolution than a steep IMF.
As first pointed out by {Tinsley} (1980), the IMF thus has
opposite effects on the luminosity and color evolution: a
more top-heavy
IMF leads to stronger luminosity evolution and weaker color
evolution.

\begin{figure*}[t]
\epsfxsize=17.5cm
\epsffile{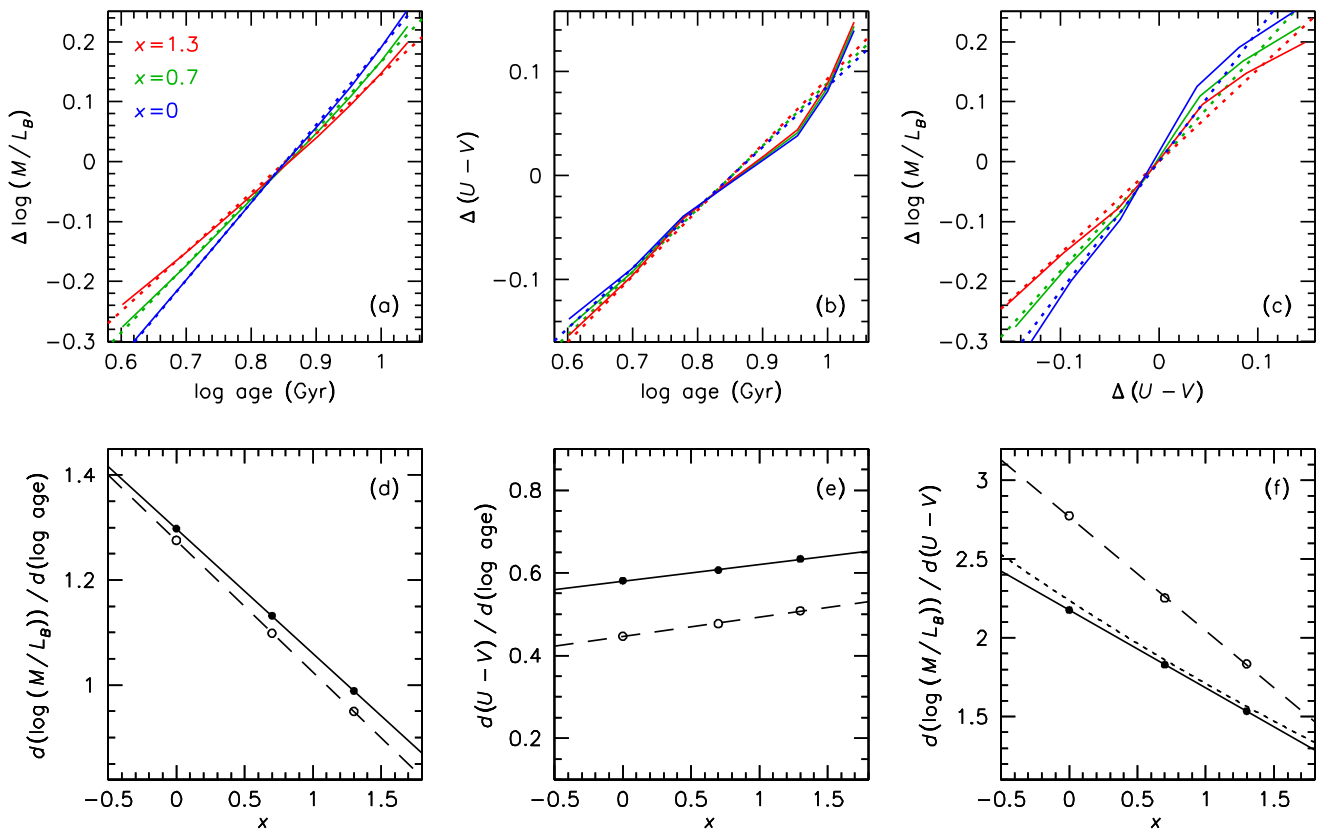}
\caption{\small
Predictions of luminosity evolution (a) and color
evolution (b) for three different
values of the logarithmic slope of the IMF $x$, from Maraston
(2005) models with [Fe/H]\,=\,0.35 and ages of $3-12$ Gyr.
Luminosity evolution is plotted as a function of color
evolution in (c).
Broken lines are power law fits to the model predictions.
Panels (d-f) show the power law coefficients obtained for the
three different values of $x$. Lines are linear fits to the
points. The dotted line in (f) is the relation implied
by the fits in (d) and (e).
Note the strong
IMF dependence of luminosity evolution and the weak
(and opposite) IMF dependence of color evolution.
Open symbols and dashed lines show results
for models with [Fe/H]\,=\,0. The color evolution, and hence
also the ratio of luminosity to color evolution, is quite
dependent on the assumed metallicity.
\label{mlcol.plot}}
\end{figure*}

\subsection{Stellar Population Synthesis Models}
\label{starpop.sec}

Following {Tinsley} (1980) and many later studies, we parameterize
luminosity and color evolution by power laws.
Power law approximations render the predicted evolution independent of the
Hubble constant and the absolute age calibration of the model, and
smooth out artifacts in the evolution of single-age
stellar populations caused by numerical effects. In our passbands
(see \S\,\ref{data.sec}) luminosity evolution takes the form
\begin{equation}
\label{mlt.eq}
\log {M}/{L_B} = \kappa_B \log (t-t_{\rm form}) + C_1
\end{equation}
and color evolution takes the form
\begin{equation}
\label{colt.eq}
U-V   = 2.5 \kappa_{U-V} \log (t-t_{\rm form}) + C_2,
\end{equation}
with $t_{\rm form}$ the luminosity-weighted mean star formation epoch
and $C_1$ and $C_2$ constants.

We use the {Maraston} (2005) stellar population synthesis
models to determine $\kappa_B$ and $\kappa_{U-V}$.
Fig.\ \ref{mlcol.plot}(a) and (b) show the evolution of the $M/L_B$
ratio and $U-V$ color in this model for 
three different values of $x$ (1.3, 0.7, 0).
These IMFs are power laws with a fixed
slope over the mass range $0.1 - 100$\,\msun.
The predictions are for a metallicity
[Z/H]\,$=0.35$. This relatively high value may be
more appropriate for massive early-type galaxies than the
Solar value
(e.g., {Worthey}, {Faber}, \&  {Gonzalez} 1992), although Solar models are
much better calibrated.
As expected, luminosity evolution
has a strong IMF dependence but color evolution only has a weak
(and opposite) IMF dependence. 
Panel (c) directly compares the evolution of the $M/L$ ratio to the
color evolution. The IMF dependence is relatively strong in this
panel, due to the opposite effects of the IMF on luminosity and color
evolution.

Broken lines in panels (a-c) are power law fits to
the model predictions. These fits generally
provide good descriptions of
the behavior of the models. The largest deviations
are for the $U-V$ colors at 9 and 12 Gyr,
where the fits are off by $+0.02$ and $-0.02$ mag
respectively. Panels (d-f) show the relation between
the powerlaw coefficients and the IMF slope. In each
case there is a well-defined relation, that can be
characterized by a simple linear fit. These fits
take the form
\begin{eqnarray}
\label{kappa1.eq}
\kappa_B   =  \frac{d(\log (M/L_B))}{d(\log (t-t_{\rm form}))} = 1.30 - 0.24 x, \\
\label{kappa2.eq}
\kappa_{U-V}  =  0.4 \frac{d(U-V)}{d(\log (t-t_{\rm form}))} = 0.23 + 0.016 x,
\end{eqnarray}
and
\begin{equation}
\frac{\kappa_B}{\kappa_{U-V}}  = 2.5 \frac{d(\log (M/L_B))}
{d(U-V)} = 5.44 - 1.23 x.
\label{kappa3.eq}
\end{equation}
The fits are shown by the solid lines in Fig.\ \ref{mlcol.plot}(d-f).
Although the fits are excellent for $0<x<1.3$ the
extrapolations outside that range are obviously
uncertain. Note that Eq.\ \ref{kappa3.eq} is a fit, and is not
identical to the combination of Eqs.\ \ref{kappa1.eq} and \ref{kappa2.eq}.
The dotted line in Fig.\ \ref{mlcol.plot}(f) shows the relation
implied by Eqs.\ \ref{kappa1.eq} and \ref{kappa2.eq}; this relation
is consistent with Eq.\ \ref{kappa3.eq} to a few percent. Open
symbols and dashed lines
are predictions for models with [Fe/H]\,=\,0. Luminosity evolution
is insensitive to [Fe/H], but color evolution has a strong
metallicity-dependence. Therefore, the relation in panel (f)
also has a strong metallicity dependence, and takes the form
\begin{equation}
\label{solkappa3.eq}
\frac{\kappa_B}{\kappa_{U-V}}  = 6.93 - 1.81 x
\end{equation}
for Solar metallicity. When determining the IMF slope in
\S\,\ref{apply.sec} results will be given both for super-Solar
models and Solar models (which
are better calibrated).

\subsection{Implementation}
\label{implem.sec}

As demonstrated in
Fig.\ \ref{mlcol.plot}, the IMF mostly influences the luminosity
evolution, and the color evolution can be viewed as an
IMF-insensitive ``clock'' constraining the age. By combining
color and luminosity evolution for a well-chosen set of
galaxies over a sufficiently large redshift range the slope
of the IMF $x$ and the formation time
of the stars $t_{\rm form}$ can, in principle,
be constrained simultaneously (although there are many
caveats -- see \S\,5).

The method requires very accurate rest-frame
color and $M/L$ measurements for a passively evolving
sample of galaxies. This is not an easy
task, as galaxy evolution is in general a complex process
involving star bursts, mergers, and morphological changes.
The most massive early-type galaxies in clusters at modest redshifts
($z<1$) come closest to
a ``purely passive'' sample.
Their colors and $M/L$ ratios have very small scatter
(e.g., {Bower}, {Lucey}, \&  {Ellis} 1992a; {J\o{}rgensen}, {Franx}, \&  {Kj\ae{}rgaard} 1996), their luminosity and color
evolution are
consistent with passive evolution of stellar populations
formed at redshifts $z\gtrsim 2$ (e.g., {van Dokkum} {et~al.} 1998a; {Holden} {et~al.} 2004), and they appear to have been largely in
place by $z\approx 0.8$ ({Holden} {et~al.} 2006, 2007).

Recently a large number of measurements of
the $M/L_B$ evolution of early-type galaxies
with masses $M>10^{11}$\,\msun\ were compiled from the literature
and placed on a consistent system ({van Dokkum} \& {van der Marel} 2007) [hereafter
vv07].
Color evolution has also been studied by many authors
(e.g., {Stanford}, {Eisenhardt}, \&  {Dickinson} 1995; {Ellis} {et~al.} 1997; {Holden} {et~al.} 2006), but not
yet for mass-limited samples and with the required accuracy.
With the availability of archival multi-color
{\em Hubble Space Telescope} (HST) WFPC2
and ACS data of a subset of the clusters in
vv07 it is now
possible to determine color and luminosity evolution of massive
early-type galaxies in clusters
simultaneously, and begin to constrain the form of the IMF at the mean
star formation epoch of massive early-type galaxies.

\section{Data}
\label{data.sec}

In this Section, data from the literature and the HST Archive
are used to determine the
evolution of rest-frame $U-V$
colors of galaxies in clusters from the vv07 sample.
Because mass measurements of individual galaxies are available
the color evolution can be determined from the zeropoint
of the color-mass (rather than the color-magnitude) relation.
Also, the same mass limit
can be applied as was applied to the $M/L$ data.
We focus on $U-V$ because redder rest-frame colors are not available
for clusters beyond $z\sim 0.5$. 

\subsection{The Color-Mass Relation in Coma}

The color-magnitude relation in the Coma
cluster was studied by {Bower} {et~al.} (1992a,b) [hereafter
BLE92].
An important goal of their study
was to determine the relative distance between the Virgo and
Coma cluster, and this required very accurate absolute color
measurements.
The data from BLE92
are combined with velocity dispersions and effective radii
from {J\o{}rgensen}, {Franx}, \&  {Kj\ae{}rgaard} (1995a,b).
Masses are calculated using
\begin{equation}
\label{massdef.eq}
\log M = 2 \log \sigma + \log r_e + 6.07,
\end{equation}
with $\sigma$ the velocity dispersion corrected to a $3\farcs 4$
diameter aperture and $r_e$ the effective radius in kpc.
Radii were converted from arcsec to kpc assuming a Hubble
flow velocity of $v_{\rm flow} = 7376 \pm 223$\,\kms\
(see vv07).

Figure \ref{comacm.plot}
shows the color-mass relation of galaxies in common between
the BLE92 color-magnitude sample and the J\o{}rgensen
et al.\ fundamental plane (FP) sample.
There is a very clear
relation, of the form
\begin{equation}
\label{colmass.eq}
U - V = 0.113 (\log M - 11) + 1.460.
\end{equation}

The slope
of the relation is determined by fitting a linear function
to all galaxies with masses $M>10^{11}$\,\msun, and the offset is
determined with the biweight statistic ({Beers}, {Flynn}, \& {Gebhardt} 1990), which
gives low weight to outliers. This relation is indicated by
the solid line in Fig.\ \ref{comacm.plot}. 
We note that the data in Fig.\ \ref{comacm.plot} are not entirely
self-consistent, as the colors were determined in a different
aperture ($11\arcsec$) than the velocity dispersions
($3\farcs 4$). This has implications for the interpretation
of Eq.\ \ref{colmass.eq}, but it has no effect on our analysis as
the distant galaxies are treated in the same way as the
Coma galaxies.

\vbox{
\begin{center}
\leavevmode
\hbox{%
\epsfxsize=8.5cm
\epsffile{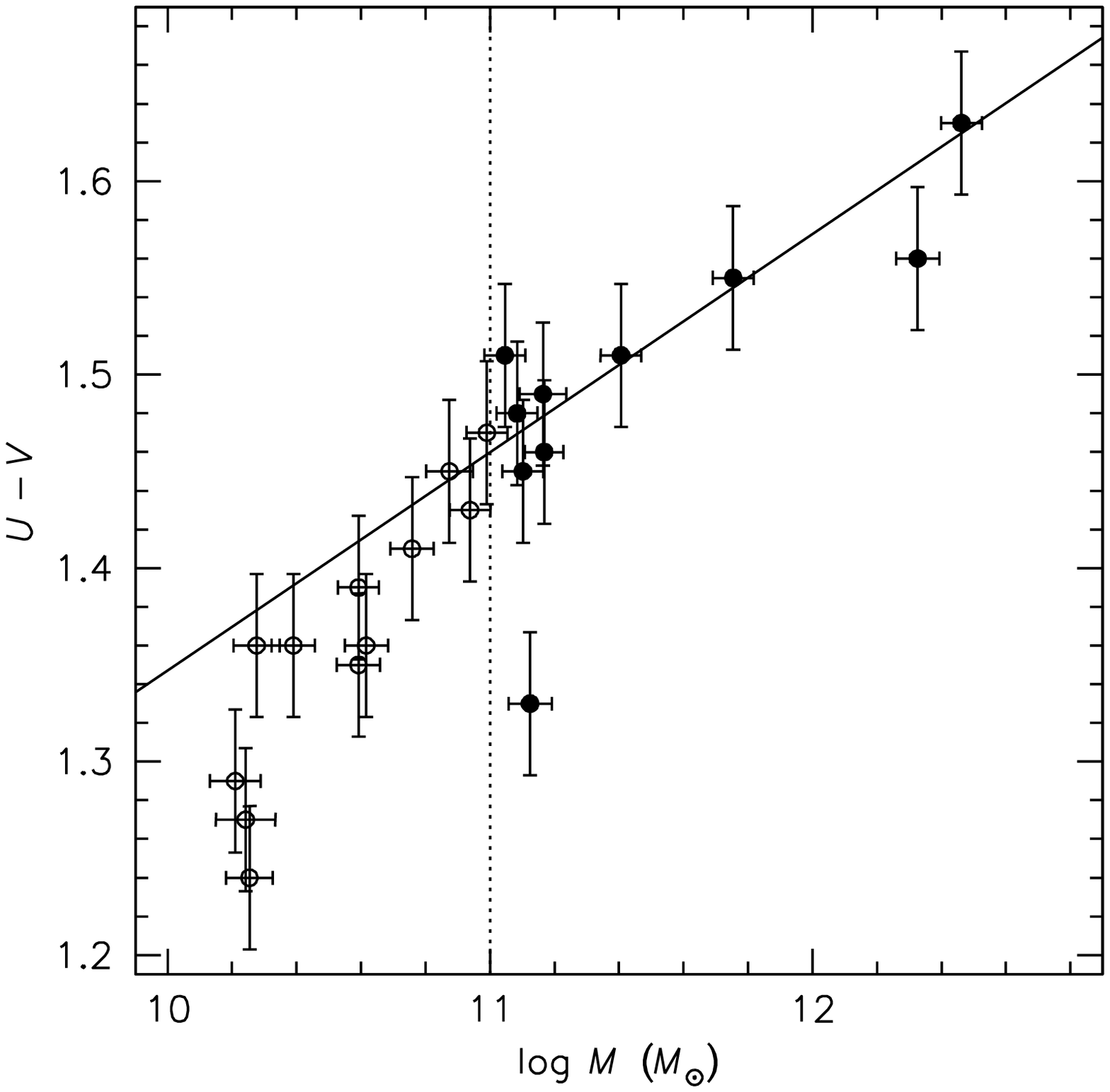}}
\figcaption{\small
Color-mass relation for the Coma cluster, using $U-V$ colors from
BLE92 and masses determined from radii and
velocity dispersions given by J\o{}rgensen et al.\ (1995a,b).
Errors in masses were determined from formal uncertainties given by
J\o{}rgensen et al. Errors in colors reflect the rms in repeat
measurements as given by BLE92.
There is a clear relation with low scatter. The line is the best
fit for galaxies with $M>10^{11}$\,\msun.
\label{comacm.plot}}
\end{center}}

The observed
scatter in the relation is very small at $\sigma_{\rm bi} = 0.029$
mag for $M>10^{11}$\,\msun. This low scatter is remarkable:
from repeat measurements of the same galaxies
BLE92 estimate that the observational
uncertainty for individual galaxies is $0.037$, which
would imply that the intrinsic scatter in $U-V$
is well below 0.03 for the most massive Coma galaxies.
However, we note that the biweight estimator gives 
very little weight to the
blue outlier at $\log M = 11.1$, and this is somewhat arbitrary given
the small sample size. The rms scatter is 0.05, similar to
values measured for clusters at higher redshift (see \S\,\ref{colevo.sec}).

In this study we are not concerned with the scatter or slope of the
color-mass relation but only with its zeropoint evolution. For each
cluster in the sample Eq.\ \ref{colmass.eq} is subtracted from the
rest-frame $U-V$ colors of early-type galaxies with $M>10^{11}$\,\msun,
and the offset $\Delta (U-V)$ is determined with the biweight estimator.
By construction, $\Delta (U-V) \equiv 0$ for the Coma cluster.
The random uncertainty in the offset due to galaxy-galaxy
color variations is 0.009.
BLE92 give a systematic error of 0.02
in the $U-V$ colors, reflecting uncertainties
in the $U$ and $V$ zeropoints and small errors due
to reddening and aperture corrections.
Adding the
random and systematic errors in quadrature gives a combined
uncertainty of 0.022. We note that the
BLE92 data are still the most accurate available for Coma,
even though they are now more than
15 years old. In fact, the comprehensive photometric
studies by {Terlevich}, {Caldwell}, \&  {Bower} (2001)
and {Eisenhardt} {et~al.} (2007) calibrate
their photometry by matching the magnitudes of
galaxies to the BLE92 values.

\subsection{Other Nearby Clusters}

No other nearby clusters are currently available with
high quality $U-V$ colors, effective radii, and velocity dispersions
for their constituent galaxies. As an example, effective
radii and velocity dispersions are available for the
clusters in the {J\o{}rgensen} {et~al.} (1996) sample, but accurate $U-V$
colors have not been measured. Conversely, {McIntosh} {et~al.} (2005)
analyzed the color-magnitude relation in the three nearby clusters
Abell 85, Abell 496, and Abell 754
using the same apertures and methods as BLE92, but
those clusters have not been the subject of FP studies.

\begin{deluxetable*}{lccccccccc}[!t]
\tabletypesize{\small}
\tablecaption{Color Evolution\label{sample.tab}}
\tablewidth{0pt}
\tablehead{ \colhead{Cluster} & \colhead{$z$} & \colhead{Camera}
& \colhead{$F_1$} & \colhead{$F_2$} & \colhead{$\alpha_{1-2}$}
& \colhead{$\beta_{1-2}$} & \colhead{$\sigma_{\rm sys}$}
& \colhead{$\Delta (U-V)$} & \colhead{$\sigma_{\rm tot}$} }
\startdata
Coma       & 0.024 & \nodata & $U$ & $V$    & \nodata & \nodata  & 0.036 &
\phantom{$-$}0.000 & 0.037 \\
Abell 2218 & 0.176 & WFPC2 & F450W & F702W & 1.230 & $-1.217$ & 0.021 &
$-0.082$  & 0.028  \\
CL\,1358+62 & 0.327 & ACS & F475W & F775W & 0.955 & $-1.253$ & 0.022 &
$-0.101$  & 0.033 \\
CL\,0024+16 & 0.391 & WFPC2 & F450W & F814W & 0.793 & $-1.181$ & 0.024 &
$-0.122$ & 0.031 \\
CL\,0016+16 & 0.546 & WFPC2 & F555W & F814W & 1.065 & $-1.227$ & 0.023 &
$-0.165$ & 0.035 \\
MS\,2053$-04$ & 0.583&WFPC2 & F606W & F814W & 1.503 & $-1.243$ & 0.035 &
$-0.167$ & 0.039 \\
MS\,1054$-03$ & 0.831 & ACS & F606W & F850LP & 0.881 & $-1.176$ & 0.033 &
$-0.200$ & 0.038 \\
RX\,J0152$-13$ & 0.837 & ACS & F625W & F850LP & 1.035 & $-1.164$ & 0.021 &
$-0.185$ & 0.024
\enddata
\end{deluxetable*}

Although  the offsets $\Delta (U-V)$ for the three
clusters of {McIntosh} {et~al.} (2005) cannot be determined directly,
they can be estimated by making use of the tight relation between
rest-frame $V$ band luminosity and mass in nearby clusters
(e.g., {J\o{}rgensen} {et~al.} 1996). Fitting directly to the BLE92 $V$
band magnitudes (in our cosmology) and {J\o{}rgensen} {et~al.} (1996)
masses gives
\begin{equation}
\label{pseudo.eq}
\log M = -0.57 M_V - 0.78,
\end{equation}
with an rms scatter of 0.14 in $\log M$. Applying this relation
to the full BLE92 sample (i.e., not just
to galaxies with measured masses) produces an offset in
the color-''pseudo''mass relation of $-0.003$, almost identical to
the offset derived from the actual color-mass relation.

Absolute $V$ band magnitudes and $U-V$ colors were taken from
Table 6 in {McIntosh} {et~al.} (2005). The $(U-V)_{3.34}$ colors
are used as they were measured using the same physical aperture
size as used by BLE92. The $V$ magnitudes are corrected to our
cosmology and converted into pseudo-masses  using Eq.\
\ref{pseudo.eq}. Offsets are determined by subtracting
Eq.\ \ref{colmass.eq} from the observed colors and determining
the biweight mean residual. The offsets are $-0.050 \pm 0.042$,
$-0.061 \pm 0.061$, and $-0.002 \pm 0.051$ for Abell 85,
Abell 496, and Abell 754 respectively.
The uncertainties are a combination of random and systematic
errors. The systematic errors
were calculated from the
listed uncertainties in the zeropoints and $k$-corrections
and a 16\,\% uncertainty in the listed reddening ({Schlegel}, {Finkbeiner}, \&  {Davis} 1998).\footnote{Note that McIntosh et al.\ list larger
systematic errors; they do not add the individual error
contributions in quadrature and also
assume a slightly larger error in the
extinction than given by Schlegel et al.\ (1998).}
All three offsets are consistent with zero, in agreement with
the analysis of {McIntosh} {et~al.} (2005) and with earlier work
by {Andreon} (2003), who studied a large sample of clusters
in the Sloan Digital Sky Survey (SDSS). The
uncertainties in $\Delta (U-V)$
are considerably larger than for Coma
due to a combination of larger photometric zeropoint uncertainties
and larger Galactic reddening toward the {McIntosh} {et~al.} (2005)
clusters.

\subsection{Data for Distant Clusters}
\subsubsection{Available HST Imaging and Observed Colors}

For each of the 14 distant clusters described
in vv07 we
determined whether HST data exist in the appropriate
filters (i.e., roughly corresponding to rest-frame $U$ and $V$).
HST data are crucial because of their photometric stability and
because the combination of seeing and color gradients
makes aperture effects difficult to control in ground-based data.
Appropriate data are available for the seven clusters listed in Table
\ref{sample.tab}.
Clusters from vv07 that are {\em not} included in the analysis
are Abell 665 ($z=0.183$), as the bluest available HST filter
corresponds to rest-frame
$V$; Abell 2390 ($z=0.228$), as the HST filters correspond
to rest-frame $B-R$;
3C\,295 ($z=0.456$), as the vv07 sample
only has one galaxy more massive than $10^{11}$\,\msun, except
3C\,295 itself; CL\,1601+42 ($z=0.539$), as HST observations were
obtained in a single filter only; and all three clusters with
$z>0.85$ as the reddest available
HST colors are much bluer than rest-frame $U-V$.

For two of the seven clusters in Table 1 accurate colors have
been published: {Blakeslee} {et~al.} (2006) provide ACS measurements of
galaxies in MS\,1054--03 ($z=0.831$) and RX\,J0152--13 ($z=0.837$).
For the five remaining clusters the HST data were obtained from the
HST archive, and reduced using standard techniques.
The WFPC2 images were
processed using a combination of STSDAS tasks and
custom scripts (see, e.g., {van Dokkum} {et~al.} 1998b).
For ACS the photometrically and astrometrically corrected
``drz'' files were obtained, and further processing was
limited to the removal of remaining cosmic rays and some
other defects. 
Galaxies in common with the vv07 samples were deconvolved
with the {\sc sclean} task in the IRAF STSDAS package, using
synthetic PSFs generated with Tiny Tim ({Krist} 1995).
The effect of deconvolution on the measured colors is
typically $\sim 0.01$ at $z\sim 0.2$ and $\sim 0.02$ at $z\sim 0.5$.

Colors were measured in apertures scaled to match an $11\arcsec$
diameter aperture at the distance of the Coma cluster, to allow
a direct comparison with BLE92. The measurements were converted
from instrumental counts to magnitudes on the Vega system using
the WFPC2 and ACS zeropoints. For WFPC2 the
zeropoints were taken from \S\,28.1 of the WFPC2 Data Handbook,
which lists values for the PC chip and each of the wide field
detectors separately. For ACS the zeropoints were taken
from the Tables on the
STScI website\footnote{http://www.stsci.edu/hst/acs/analysis/zeropoints/},
which include small corrections made after publication of
{Sirianni} {et~al.} (2005). For the two $z\approx 0.83$ clusters
the colors were taken from Tables 1 and 2 of {Blakeslee} {et~al.} (2006).
A small aperture correction was applied to each galaxy,  as
Blakeslee et al.\ used apertures containing 50\,\% of the light
rather than an aperture of fixed size (see Appendix A). The
colors were transformed from the AB to the Vega system using
the AB offsets listed in {Sirianni} {et~al.} (2005).

\subsubsection{Rest-frame Colors and Systematic Errors}
\label{syserrors.sec}

A key step in the analysis is the transformation of the observed
colors in the various HST filters to rest-frame $U-V$.
Following {van Dokkum} \& {Franx} (1996) and many
later studies (e.g., {Blakeslee} {et~al.} 2006) we derive
transformations of the form
\begin{eqnarray}
U_z & = & F_2 + \alpha_1 (F_1 - F_2) + \beta_1 + 2.5 \log (1+z) \\
V_z & = & F_2 + \alpha_2 (F_1 - F_2) + \beta_2 + 2.5 \log (1+z)
\end{eqnarray}
with $U_z$ the rest-frame $U$ band, $V_z$ rest-frame $V$, $F_1$ and
$F_2$ the observed HST filters (listed in Table 1), and
$\alpha$ and $\beta$ determined from template spectra.
The rest-frame color is given by
\begin{eqnarray}
\label{coltrafo.eq}
(U-V)_z & = & (\alpha_1 - \alpha_2) (F_1-F_2) + (\beta_1 - \beta_2)
\nonumber\\
  & = & \alpha_{1-2} (F_1 - F_2) + \beta_{1-2}
\end{eqnarray}
Unless $F_1$, $F_2$ correspond exactly to $U_z$, $V_z$
the values of $\alpha$ and $\beta$ have a (modest)
dependence on the assumed spectral type.
For each cluster the appropriate values
were determined in the following way. We used synthetic
templates from  {Yi} {et~al.} (2003) with a large range of ages to
establish the relation between the coefficients and the
observed color in the HST filters. The {Yi} {et~al.} (2003) templates
were chosen because they may be better calibrated in the
rest-frame ultra-violet than the {Bruzual} \& {Charlot} (2003) models; at
rest-frame wavelengths $\lambda \gtrsim 3600$\,\AA\ the
two sets of models are identical. Typically,
$\alpha$ and $\beta$ vary by less than a few percent
for ages varying from 1 -- 12 Gyr; the amount of variation
depends largely on how well the observed HST filters
match the rest-frame $U$ and $V$ bands. The relations
between $\alpha$, $\beta$ and the observed colors
$F_1-F_2$ were fitted with smooth
functions, and the values corresponding to the median observed
color of galaxies in the sample were adopted. Values
of $\alpha_{1-2} \equiv \alpha_1 - \alpha_2$ and
$\beta_{1-2} \equiv \beta_1 - \beta_2$ are listed in
Table 1.

\begin{figure*}[t]
\epsfxsize=13.3cm
\epsffile[-20 187 423 654]{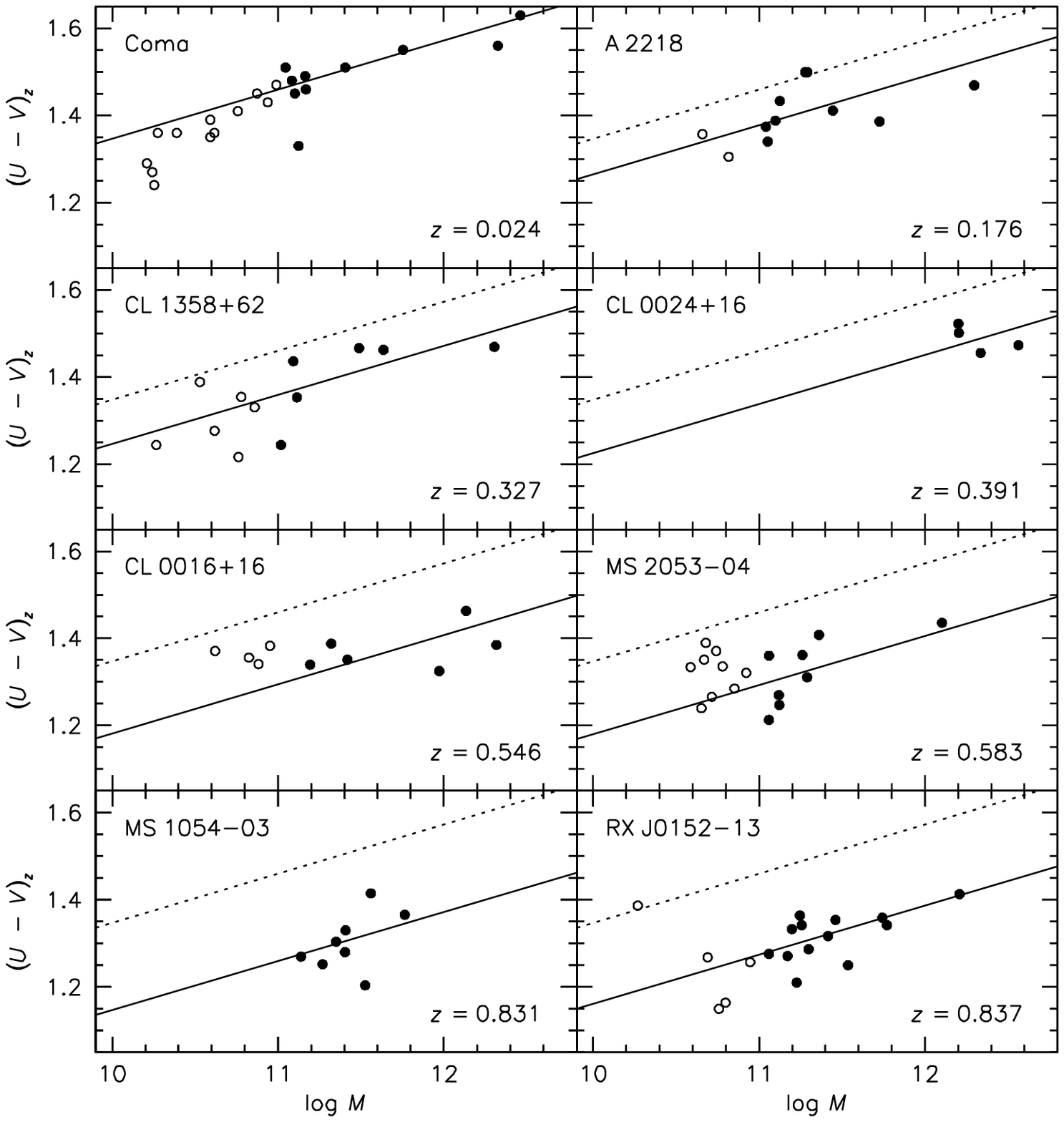}
\caption{\small
Evolution of the color-mass relation, as determined from HST photometry
in combination with ground-based dynamical measurements. Solid symbols
indicate galaxies with dynamical masses $>10^{11}$\,\msun, which 
are used for determining the evolution of the relation. The broken line
in each panel is the best fit to the Coma cluster. Solid lines are
the best fitting relations for each cluster,
keeping the slope fixed but allowing the zeropoint to vary.
The data confirm that massive cluster
galaxies gradually redden with time, as
expected from stellar evolution.
\label{colmass.plot}}
\end{figure*}

The rest-frame $U-V$ colors
are subject to several sources of systematic uncertainty.
The zeropoints of the HST filters are accurate to
$\lesssim 0.02$ mag. The dominant source of
zeropoint error is thought to be the absolute calibration
of Vega; e.g., updates to the ACS zeropoints in 2006 due to a
recalibration of Vega were $\lesssim 0.017$. As we are
always dividing the flux in two ACS or WFPC2 passbands, such
errors (nearly) cancel. Other effects that plague absolute
calibration of (particularly) WFPC2 photometry, such as
variations in charge transfer efficiency and the long-short
anomaly, also cancel when measuring colors. Therefore, we
estimate that the systematic uncertainty
of the measured colors (before
correcting for reddening) is 0.010 for WFPC2 and 0.015
for ACS (as it has a shorter calibration history).
Uncertainties in the Galactic reddening corrections are
$\approx 16$\,\% of the applied values ({Schlegel} {et~al.} 1998),
which is typically $\approx 0.01$ mag for our clusters.
This error is largest for CL\,0024+16, at 0.021.

The transformations to rest-frame colors also
have some uncertainty, particularly if the observed HST
filters are not well matched to rest-frame $U-V$.
This uncertainty was assessed by varying the templates
used for the transformation, comparing results for
the {Yi} {et~al.} (2003) templates, {Bruzual} \& {Charlot} (2003) templates,
and empirical templates from {Coleman}, {Wu}, \& {Weedman} (1980).
For most clusters the error is $\sim 0.015$ mag. For
MS\,2053--04 and MS\,1054--03, which have the
worst match of observed and rest-frame filters among
the seven clusters,  the uncertainties are
0.025 and 0.030 respectively. We note that, once again,
a potentially important systematic effect cancels: the form of
the synthetic rest-frame $U$ and $V$ filters, and their
AB zeropoints, is the same for all distant clusters, which
means that the distant clusters are all on the exact same
synthetic photometric system. There may be a systematic
offset between this system and Coma, although we use
the same {Bessell} (1990) passbands as {Landolt} (1992).
This possible offset was taken into account by adding a 0.03
mag error in quadrature to the systematic uncertainty
of the Coma offset. Assuming that the sources of systematic
error are independent the combined error is given by
\begin{equation}
\sigma_{\rm sys}^2 = \alpha_{1-2}^2 (\sigma_{\rm zero}^2 + \sigma_{\rm red}^2)
+ \sigma_{\rm trafo}^2.
\end{equation}
The values of $\sigma_{\rm sys}$ are listed in Table 1.
They are generally small, of order 2\,\% -- 3\,\%.
The reliability of these numbers is empirically assessed
below.

\subsection{Evolution of the Color-Mass Relation}
\label{colevo.sec}

For each of the distant clusters in the sample the color-mass
relation is constructed by combining the rest-frame $U-V$
colors with masses determined using Eq.\ \ref{massdef.eq}.
The required effective radii and velocity dispersions are
taken from the literature and corrected to a consistent
system (see Appendix A of vv07). The color-mass relations
are shown in Fig.\ \ref{colmass.plot}. Galaxies at higher
redshift have bluer rest-frame $U-V$ colors, as expected from
stellar evolution and from many previous studies of color
evolution
(e.g., {Stanford} {et~al.} 1995, 1998;
{Kodama} {et~al.} 1998; {Holden} {et~al.} 2004).


The evolution of the zeropoint is quantified by subtracting
Eq.\ \ref{colmass.eq} from the datapoints, and determining the
center of the distribution of residuals with the biweight
statistic. Choosing the average or the median rather than
the biweight slightly changes
the offset of each cluster (always well within its error), but
does not have an effect on our conclusions.
We verified
that the slopes for the high-redshift clusters are consistent
with the Coma slope by fitting the color-mass relation of all
galaxies in the seven distant clusters simultaneously,
after removing the zeropoint evolution
derived below. The difference between this slope and the adopted
Coma slope is only $0.01 \pm 0.02$. Varying the adopted slope
within these limits does not change the derived offsets.
The analysis relies on the
assumption that the slopes do not vary appreciably from cluster
to cluster. The samples of galaxies with mass measurements are
too small to determine the slopes of the color-mass
relations in individual clusters, but we
note that Holden et al.\ (2004) find no evidence for either
evolution or cluster-to-cluster variation in the slope of the
color-magnitude relation.

The offsets are listed
in Table 1. For each cluster the uncertainty in the offset $\sigma_{\rm tot}$
is the quadratic sum
of the systematic error and the
uncertainty in the biweight mean (which is caused by the galaxy-galaxy
scatter within each cluster). For most clusters the random
uncertainty is of the same order as the systematic uncertainty.
The evolution of the zeropoint of the color-mass relation is
shown in Fig.\ \ref{colevo.plot}. The evolution is very regular
for these clusters, which is perhaps not surprising given previous
studies (e.g., {Kodama} {et~al.} 1998; {Andreon} 2003; {Holden} {et~al.} 2006) and
the small cluster-to-cluster scatter in the evolution of
$M/L$ ratios (see, e.g., vv07).

The solid line in Fig.\ \ref{colevo.plot} is the best fitting linear
function to the data (including the three Mcintosh et al.\ [2005]
clusters):
\begin{equation}
\label{colevo.eq}
\Delta (U-V) = (-0.028  \pm 0.018) - (0.214 \pm 0.041) z.
\end{equation}
The observed cluster-to-cluster scatter around this line is very small at
$0.022 \pm 0.005$ for all 11 clusters or $0.019 \pm 0.005$ for the eight
clusters with direct mass measurements. This remarkably small
cluster-to-cluster scatter is
qualitatively consistent with previous work by {Andreon} (2003),
although we note that our study uses a much smaller number of clusters.

The small measured scatter provides an external
check on the reliability of the systematic errors derived in
\S\,\ref{data.sec}. Focusing on the eight clusters with direct
mass measurements, the expected scatter from measurement errors
alone is 0.032, higher than the observed scatter of 0.019. Assuming
the errors are correct the probability
of measuring a scatter $\leq 0.019$ is only 7\,\%, which means
the systematic
errors are much more likely to be overestimated than underestimated.
A formal limit on the systematic error can be derived by requiring
that the observed scatter of 0.019 has a one-sided probability
$>5$\,\%, and parameterizing the total error for each cluster
by $\sigma_{\rm tot}^2 = \sigma_{\rm random}^2 + F^2 \sigma_{\rm sys}^2$,
with $F$ the same for all clusters. From Monte-Carlo simulations
we find that $0\leq F \leq 1.12$, i.e., the systematic errors are
consistent with zero and underestimated by at most $12$\,\%.


Most studies of the color-magnitude relation at high redshift have
focused on the galaxy-galaxy scatter in the relation.
Although this is not a focus of the
present paper, we can robustly measure the scatter for
galaxies with $M>10^{11}$\,\msun\ by removing the average trend
(Eq.\ \ref{colevo.eq}) and combining the galaxies in 
all distant clusters.
The observed biweight scatter in the full sample $\sigma_{U-V}
=0.056 \pm 0.005$.
The scatter expected from observational error is less than
$\sim 0.02$, and we estimate that the intrinsic scatter is $\approx 0.052$.
Averaging the three clusters at $0.17<z<0.40$, the two clusters at
$0.54<z<0.59$, and the two clusters at $0.83<z<0.84$, we find no
evidence for evolution with redshift: the scatter in each redshift
bin is consistent with the value for the full sample. 
The value for the scatter that we find is significantly lower
than results from previous studies 
(e.g., {Holden} {et~al.} 2004; {Blakeslee} {et~al.} 2006). This is most likely due
to a dependence of the scatter on galaxy mass. We also note that
the color-mass relation is expected to have a smaller
scatter than the color-luminosity relation even within the
same sample of galaxies, if the scatter is
due to age variations (see {van Dokkum} {et~al.} 1998b).

\vbox{
\begin{center}
\leavevmode
\hbox{%
\epsfxsize=8.5cm
\epsffile{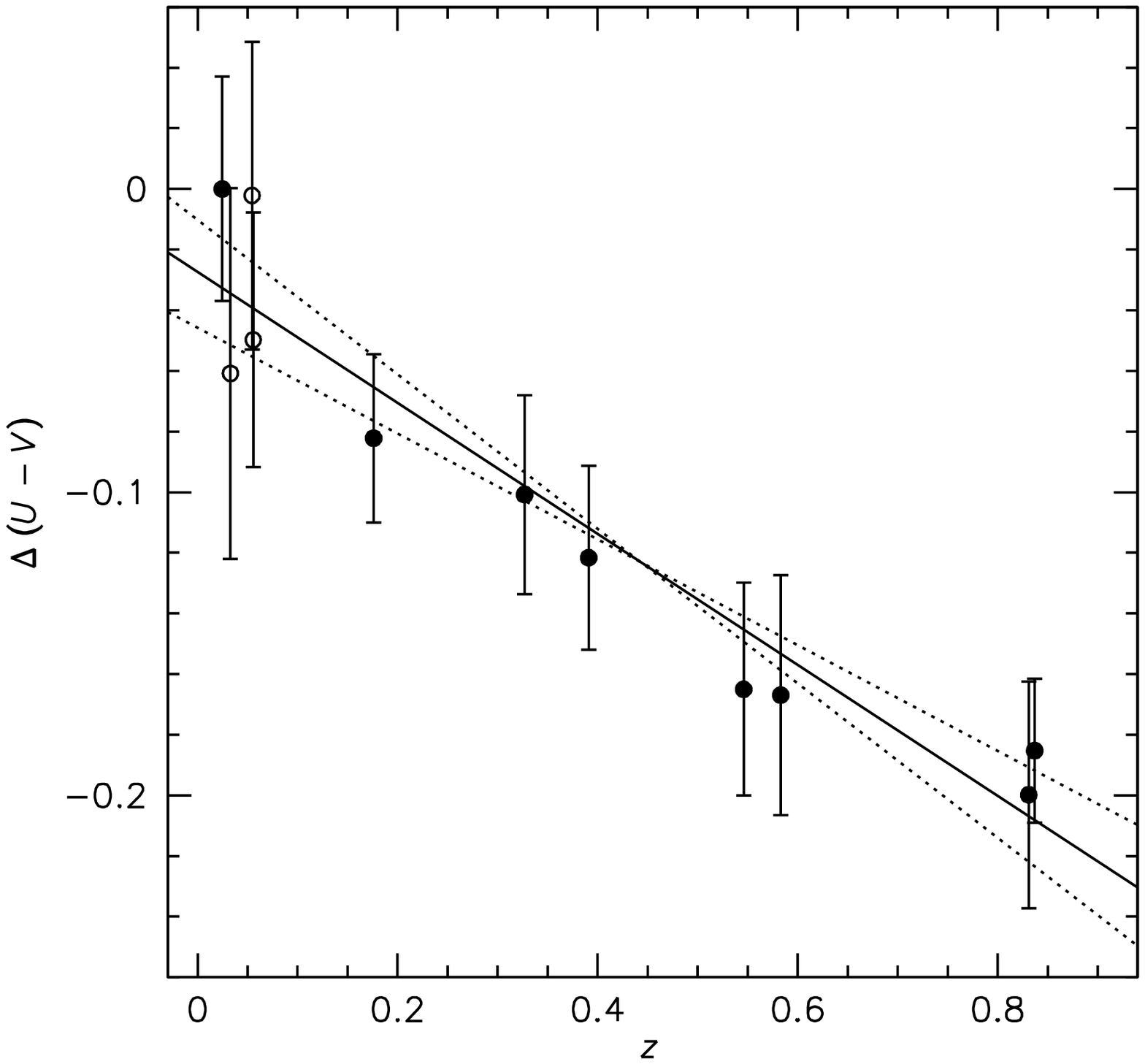}}
\figcaption{\small
Evolution of the zeropoint of the color-mass relation.
Open
symbols are three clusters from McIntosh et al.\ (2005), for which
no direct mass measurements are available. Note that there
may be a small systematic offset between the Coma cluster
at $z=0.024$ and the other clusters due to differences in
methodology. The solid line is
the best-fit linear function to the data; broken lines indicate
the $\pm 1 \sigma$ uncertainty in the slope of the relation.
\label{colevo.plot}}
\end{center}}

\section{Fitting}
\label{apply.sec}

\subsection{Constraints on the IMF}
\label{direct.sec}

As discussed in \S\,\ref{method.sec} luminosity and color
evolution each depend on the IMF and on the age of the stellar population,
but the age-dependence drops out when comparing the amount of
luminosity evolution to the amount of color evolution.
Figure \ref{direct.plot} shows the measured evolution in $\log (M/L_B)$
as a function of the evolution in $U-V$. This Figure is the
equivalent of Fig.\ \ref{mlcol.plot}(c). The values for
$\Delta \log (M/L_B)$ are taken directly from vv07, and the
values for $\Delta (U-V)$ are listed in Table 1.

As expected, there is a clear relation, with galaxies becoming both
bluer and more luminous at earlier times. The solid line is a linear
fit to the data (taking the errors in both parameters into account)
of the form
\begin{equation}
\label{fit.eq}
\Delta \log (M/L_B) = 
2.8^{+0.7}_{-0.5} \times \Delta (U-V)
\end{equation}
The residuals from this fit are consistent with expectations
from the uncertainties in the datapoints. The solid and dashed
red lines in
Fig.\ \ref{fit.eq} are predictions for a Salpeter-like\footnote{The
observations constrain the IMF in the mass range around 1\,\msun, and
at those masses differences between the {Salpeter} (1955),
{Kroupa} (2001), and {Chabrier} (2003) IMFs are very small.
In vv07 it is shown explicitly that these various IMFs all predict
very similar $M/L$ evolution over the relevant range of ages; hence
the term ``Salpeter-like'' or ``standard'' to denote IMFs with slope
$x\approx 1.3$ at $m\gtrsim 1$\,\msun.}
IMF with $x=1.3$ and two different metallicities.
Remarkably, the observed relation is much steeper than expected
from a standard IMF, even for Solar metallicity. For a given
color evolution the luminosity evolves faster than expected,
indicating a top-heavy IMF with a relatively large fraction
of rapidly evolving massive stars.

The logarithmic slope of the IMF 
follows directly from Eqs.\ \ref{kappa3.eq},
\ref{solkappa3.eq}, and \ref{fit.eq}. We find
$x = -1.4^{+1.1}_{-1.6}$ for super-Solar metallicity and
$x=-0.1^{+0.8}_{-1.1}$ for Solar metallicity, where the
uncertainties reflect the 68\,\% confidence interval. Both
values are well below
the canonical Salpeter value of $x=1.3$. Negative values
of $x$ have a large systematic uncertainty, as they require
significant extrapolation of the {Maraston} (2005)
models. Therefore, the results can best be expressed
as upper limits, particularly for the super-Solar metallicity
model. The 90\,\% confidence upper limits are $x<0.1$
for the super-Solar model and $x<0.9$ for
Solar metallicity.
The Salpeter value can be ruled out at
$>98$\,\% confidence.

\vbox{
\begin{center}
\leavevmode
\hbox{%
\epsfxsize=8.5cm
\epsffile{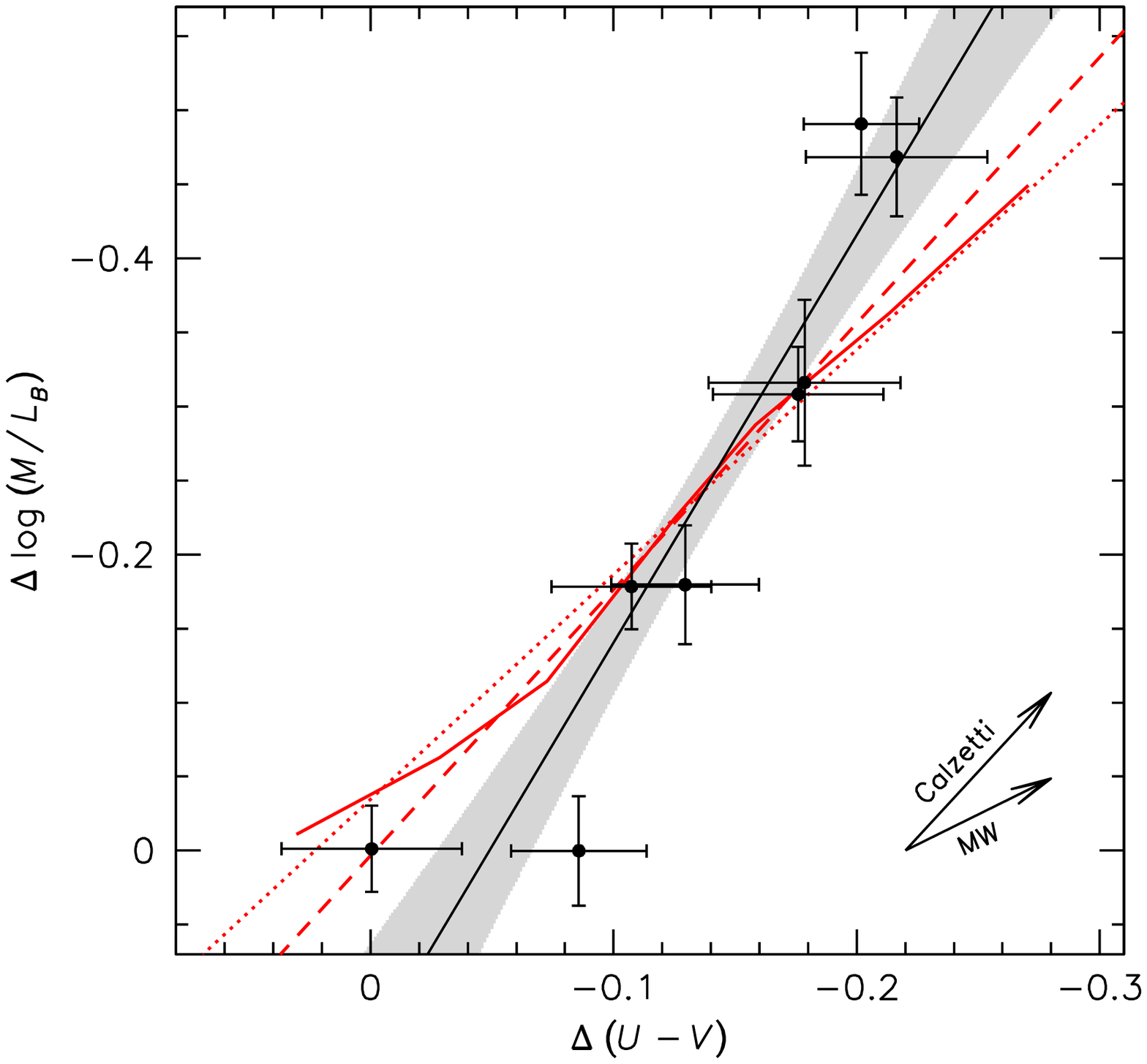}}
\figcaption{\small
Comparison of luminosity evolution to color evolution, for
massive galaxies in clusters at $0.02 \leq z \leq 0.83$.
The point at (0,0) is the Coma cluster.
Red lines show the expected relation for IMFs with a slope
of $x=1.3$ near 1\,\msun. The solid line is a
Maraston (2005) model of super-Solar metallicity; the broken
line is the best-fitting powerlaw to this model.
The dashed
line represents a Maraston (2005) model with Solar metallicity.
The solid black line shows the best-fitting
relation, with the $1\sigma$ uncertainty in grey. The evolution of
the $M/L$ ratio is faster than expected for ``standard'' IMFs,
and is consistent with expectations for a top-heavy IMF.
Arrows indicate the corrections that would have to be applied
to account for redshift-dependent dust effects.
\label{direct.plot}}
\end{center}}

\begin{figure*}[t]
\epsfxsize=17.5cm
\epsffile{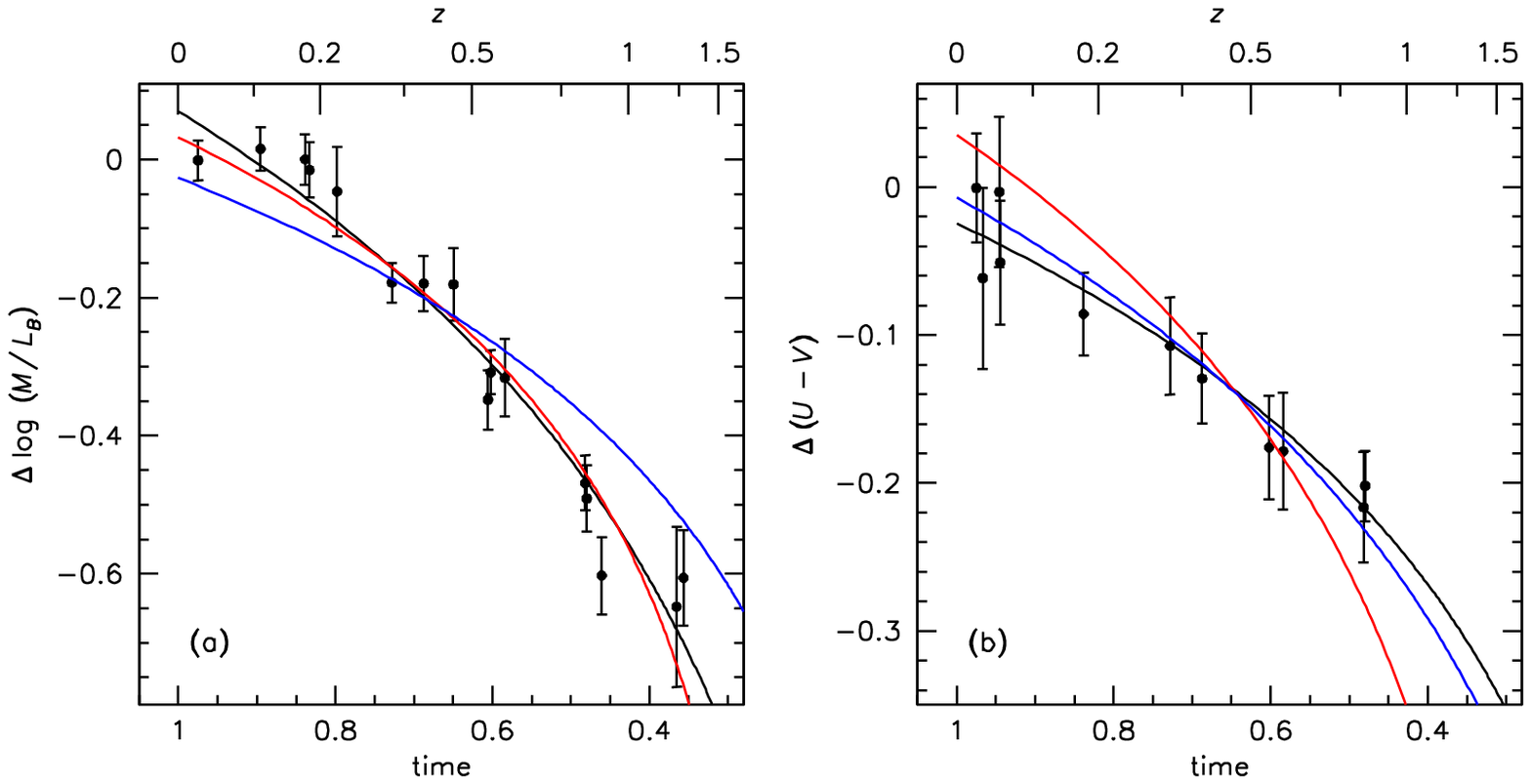}
\caption{\small
Redshift evolution of the rest-frame $M/L_B$ ratio of massive cluster
galaxies from vv07 (a), and of the rest-frame $U-V$ color (b).
Corrections of $-0.05 z$ and $-0.023 z$ have been applied to the
data in panels (a) and (b) respectively, to account for (mild)
progenitor bias. The lines show models with different formation
redshifts $z_{\rm form}$ and IMF slopes $x$. The red model has
a standard IMF and $z_{\rm form}=2$, the blue model has
a standard IMF and $z_{\rm form}=6$, and the black model
has a flat IMF and $z_{\rm form}=6$. Only the black model is a good
fit to both datasets.
\label{fitboth.plot}}
\end{figure*}

\subsection{Joint Constraints on the IMF and Formation Epoch}
\label{joint.sec}

Directly fitting $\Delta \log (M/L_B)$ as a function of
$\Delta (U-V)$, as done in \S\,\ref{direct.sec}, has several advantages:
the same galaxies are used to measure the relevant parameters,
limiting selection effects; redshift-dependent selection effects
such as progenitor bias cancel; and the IMF is the only free
parameter (at fixed metallicity), as the formation epoch of
the stars $z_{\rm form}$ also cancels. However, constraining
$z_{\rm form}$ better is of great interest in its own right, and
allows us to associate a particular time in the history of the
universe with the IMF result.
Here we fit the redshift evolution of the $M/L_B$ ratio and the
$U-V$ color simultaneously, and derive joint constraints on the
slope of the IMF $x$ and the star formation epoch $z_{\rm form}$.
The goals are to determine the luminosity-weighted star formation
epoch of massive cluster galaxies
in a self-consistent way and to verify the results
of \S\,\ref{direct.sec}.
Although the additional parameter implies more freedom in the fits
this is compensated by the fact that more data can be used, as we are
no longer restricted to the sample of eight clusters which have both
$M/L$ and color information.

The data are shown in Fig.\ \ref{fitboth.plot}. Panel (a) shows
offsets in $\Delta \log (M/L_B)$
from vv07, and panel (b) shows the offsets in $U-V$ determined
in the present study.
The dataset is more extensive than used in the analysis of
\S\,\ref{direct.sec}. The $M/L$ sample includes data from the
SDSS and for seven additional distant clusters. The color
sample includes the three clusters from {McIntosh} {et~al.} (2005),
which help constrain the low redshift end.
Small corrections for progenitor bias have
been applied, following vv07: $-0.05 z$ for $\Delta \log (M/L_B)$ and
$-0.023$ for $\Delta (U-V)$. The color correction was chosen to
be consistent with the correction for luminosity evolution and
with Eq.\ \ref{fit.eq}. As shown in \S\,\ref{prog.sec},
setting the progenitor bias
to zero leads to slightly higher formation redshifts
but has virtually no impact on the IMF constraints.

The data are fit by creating models over a grid of $x$ and
$t_{\rm form}$, with $x$ the logarithmic slope of the IMF and $t_{\rm form}$ the
mean luminosity weighted formation time of the stars. For
each value of $x$ corresponding values of $\kappa_B$ and
$\kappa_{U-V}$ are determined using Eqs.\ \ref{kappa1.eq}
and \ref{kappa2.eq}. Next, Eqs.\
\ref{mlt.eq} and \ref{colt.eq} are used to determine the
expected evolution with cosmic time 
for each combination
of $x$ and $t_{\rm form}$. The
$\chi^2$ values of the fits are then calculated by
\begin{eqnarray}
\chi^2_{x, t_{\rm form}} =
\sum_{i=1}^{16}
\left(\frac{(\Delta \log M/L_B)_{\rm obs}^i-(\Delta \log M/L_B)_{x,t_{\rm form}}^i}
{\sigma_{\Delta \log M/L_B}^i}
\right)^2 + \nonumber \\
\sum_{j=1}^{11}
\left(\frac{\Delta (U-V)_{\rm obs}^j-\Delta (U-V)_{x,t_{\rm form}}^j}
{\sigma_{\Delta(U-V)}^j}
\right)^2 
\end{eqnarray}
The fit has four free parameters: $x$, $t_{\rm form}$ (or
equivalently $z_{\rm form}$), and the
normalizations to the $M/L$ and color data.
Figure \ref{contour.plot}
shows the 68\,\%, 95\,\%, and 99\,\% confidence intervals in the
$x$, $t_{\rm form}$ plane, as determined from the $\Delta \chi^2$ values
appropriate for two interesting parameters. Solid contours are
for super-Solar metallicity ([Fe/H]\,=\,0.35) {Maraston} (2005)
models and broken contours are for Solar metallicity models.
Flat IMFs and high
formation redshifts are clearly preferred, for both
metallicities.

\vbox{
\begin{center}
\leavevmode
\hbox{%
\epsfxsize=7.7cm
\epsffile{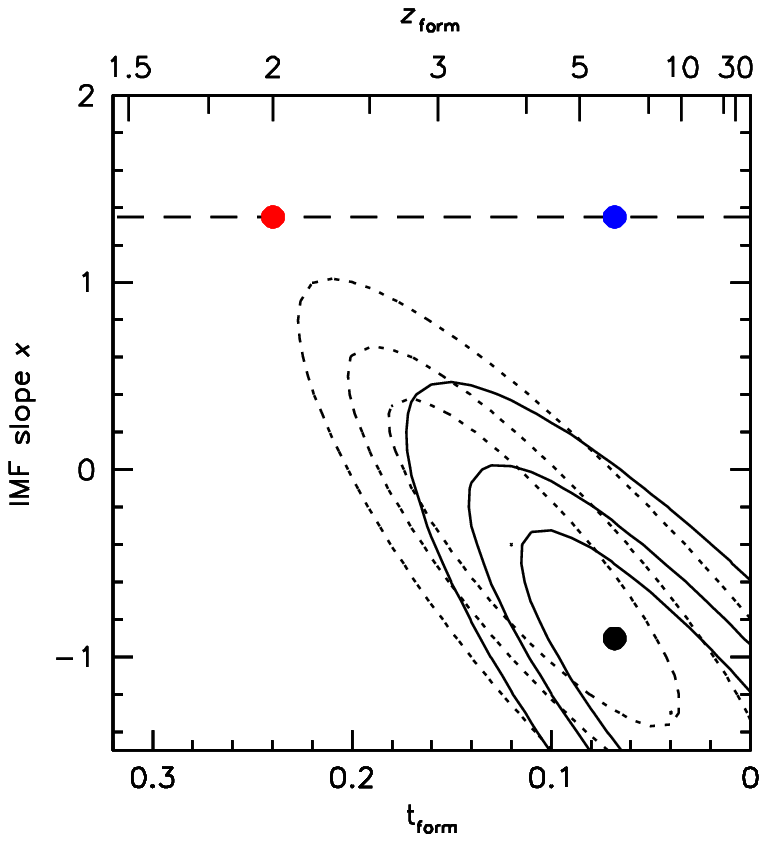}}
\figcaption{\small
Results from simultaneous fits to the redshift evolution of
$\log M/L_B$ and $U-V$ color. Contours indicate 68\,\%,
95\,\%, and 99\,\% confidence limits, as determined from
the $\Delta \chi^2(t_{\rm form},x)$ distribution. Solid contours
are for Maraston et al.\ (2005)
models with [Fe/H]\,=\,0.35, and broken contours are for
[Fe/H]\,=\,0. Early star formation and top-heavy IMFs are clearly
preferred. Red, blue, and black dots indicate the locations
of the models shown in Fig.\ \ref{fitboth.plot}.
\label{contour.plot}}
\end{center}}

The red, blue, and black dots indicate examples
of models with different parameters. The fits to the
luminosity and color evolution for these three models are
shown in Fig.\ \ref{fitboth.plot}, and illustrate how
the data simultaneously constrain $x$ and $t_{\rm form}$.
The red model has a standard Salpeter IMF and a low formation
redshift $z_{\rm form}=2$. This model provides an excellent
fit to the evolution of the $M/L_B$ ratio, as shown in
Fig.\ \ref{fitboth.plot}(a) and discussed in vv07.
However, this model fits the data in Fig.\ \ref{fitboth.plot}(b)
poorly: the observed color evolution is much slower than predicted by
this young model. The blue model is a Salpeter model with
a high formation redshift $z_{\rm form}=6$. This model
fits the slow color evolution very well, but underpredicts
the $M/L$ evolution in Fig.\ \ref{fitboth.plot}(a).
The only  models that fit the data in both
panels simultaneously
have high formation redshifts (to fit the colors)
and a flat IMF (to fit the $M/L$ ratios). The black model
has $z_{\rm form}=6$ and $x=-0.9$. This model provides an
excellent fit to the data in both panels.

The formal best-fitting values are $z_{\rm form}=7.6$,
$x=-1.2$ for super-Solar metallicity, and $z_{\rm form}=3.7$,
$x=-0.3$ for Solar metallicity. The best-fit values for the
IMF slope are in very good agreement with those derived in
\S\,\ref{direct.sec}. As discussed earlier negative values
for $x$ have large systematic
uncertainties, and may not be very meaningful. The
super-Solar solutions are therefore somewhat difficult
to interpret. Formal 68\,\%
confidence intervals for the Solar metallicity solution
are $-1.0\leq x\leq 0.1$ and $2.9 \leq z_{\rm form} \leq 6.0$.
These results are consistent with the analysis of the
$M/L$ evolution alone in vv07, where it was noted
that the best-fitting
$z_{\rm form} \approx 4$ for top-heavy IMFs with
$x=0$.

\section{Caveats and Uncertainties}
\label{caveats.sec}

Although straightforward in principle,
the IMF test discussed in this paper has several
associated uncertainties, some having
to do with the data and some with the models that are used
to fit the data.

\subsection{Effects of Sample Selection and Progenitor Bias}
\label{prog.sec}

Every cluster for which
the required data were available in the literature and/or the
HST archive was included in the analysis, and as
they come from a wide range of projects
by many different research groups this may have introduced
systematic errors. In particular, it may be that the sample
of clusters with FP measurements is somehow different from
the sample of clusters with color measurements. This potential
bias is explicitly addressed in \S\,\ref{direct.sec}, where
the analysis was limited to
clusters in common between the two samples.
As a further test, the analysis in \S\,\ref{joint.sec}
was repeated after removing the three clusters with $z>0.83$
from the sample. These clusters have a large weight in the
$M/L$ analysis as they are at the highest redshifts, and
they have no counterparts in the color sample. The effects
are small, but not negligible: for the super-Solar
model the best-fitting IMF slope changes from $-1.2$ to
$-1.0$ and
the best fitting formation redshift changes from $7.6$ to
$5.9$ when the highest redshift clusters are removed.

Redshift-dependent selection effects such
as progenitor bias are not a major concern in this particular
study. If we systematically miss the youngest progenitors
of today's clusters, or the youngest progenitors of today's
early-type galaxies within clusters, the formation
redshifts that we derive may be too high
but the IMF results should be robust. The IMF is constrained
by the ratio of the amount of luminosity evolution to
the amount of color evolution, and progenitor bias would
(to first order) equally affect the colors and luminosities.
This can be tested explicitly by varying the progenitor bias
correction to the $M/L$ ratios and colors, and repeating the
analysis. 
Increasing progenitor bias by a factor of two gives the same
IMF slope ($-1.2$) as the model adopted in \S\,\ref{joint.sec},
and a lower formation redshift (4.8 rather than 7.6).
Setting the progenitor bias to zero
again gives the same IMF slope ($-1.2$), and a formal
best-fitting formation redshift $z_{\rm form}=\infty$.

\subsection{Systematic Errors in the Photometry}

As discussed in \S\,\ref{syserrors.sec} the low cluster-to-cluster
scatter in the color evolution implies that
the errors in the photometry
are more likely overestimated than underestimated. Nevertheless,
absolute photometry is challenging, and it is difficult to rule
out subtle systematics in the various corrections and
transformations that have been applied to the data.
Increasing the errorbars on the colors by 50\,\% (much more than
is plausible -- see \S\,\ref{syserrors.sec}) obviously
loosens the constraints on the IMF, but not by a large amount.
Repeating the analysis of \S\,\ref{direct.sec} changes the
upper limits on the IMF slope from $x<0.1$ (for super-Solar
metallicity) and $x<0.9$ (for Solar metallicity) to
$x<0.4$ and $x<1.1$ respectively.

Aside from general concern about the assumed errors,
a specific worry
is the tie between the low redshift data and
the (homogeneous) synthetic photometric system
that is used for the distant clusters.
Absolute photometry in the $U$ band is particularly difficult:
the bandpass is strongly influenced by the detector and the
atmosphere, and the spectral energy distributions
of old galaxies have a very steep slope around 3800\,\AA.
We dealt with this issue in \S\,\ref{syserrors.sec} by adding
a 3\,\%  systematic uncertainty
in the Coma offset. A more conservative approach is to simply
discard Coma altogether. Interestingly, removing Coma
has the effect of tightening the
limits on the IMF slope. As an example, for Solar
metallicity the 90\,\% upper limit changes from 0.9 to
0.3. The reason for this is readily apparent from Fig.\ \ref{direct.plot}:
Coma is the point with the largest deviation from the best-fitting
line, and the relation becomes much steeper when it is removed.\footnote{Abell
2218 is the point with the largest deviation in the opposite
sense to Coma; arbitrarily
removing this cluster (while retaining Coma)
changes the 90\,\% upper limit from 0.9 to 1.3 for Solar
metallicity.} We note that the mean $M/L_B$ ratio of
Coma galaxies may also be somewhat anomalous, as it
deviates from the best-fitting
relation in Fig.\ \ref{fitboth.plot}a. Adopting the $M/L_B$
ratio from this fit rather than the actual value brings the
cluster in close agreement with the solid line in Fig.\ \ref{direct.plot}.

\subsection{Interpretation of the Fundamental Plane and $M/L$ Ratios}
\label{whatml.sec}

The evolution of the mean
$M/L$ ratio, shown in Fig.\ \ref{fitboth.plot}(a),
is derived from the evolution of the fundamental plane (FP)
relation ({Djorgovski} \& {Davis} 1987). As discussed in detail
in several papers (e.g., {Franx} 1993; {van Dokkum} \& {Franx} 1996; {Treu} {et~al.} 2001)
the empirical FP relation can be rewritten as a relation between
$M/L$ ratio and mass, if it is assumed that $M \propto \sigma^2 r_e$
(with $\sigma$ the stellar velocity dispersion and $r_e$ the
effective radius) and early-type galaxies are a homologous family.
If these assumptions are valid the observed evolution of the FP should track
the evolution of the zeropoint
of the underlying $M/L$ -- mass relation.
However, this may not be
the case if galaxies undergo significant structural changes over time,
due to merging or other processes 
(see, e.g., {Almeida}, {Baugh}, \& {Lacey} 2007).

Strong gravitational
lenses provide an independent check on the FP interpretation.
Current results indicate that
the evolution of the $M/L$ ratio as derived from strong lenses is
consistent with that derived for field early-type galaxies
({Rusin} \& {Kochanek} 2005), although the systematic
uncertainties in this comparison are still fairly substantial
(see, e.g., {Treu} \& {Koopmans} 2004; {van der Wel} {et~al.} 2005; {Rusin} \& {Kochanek} 2005).
Recently {van der Marel} \& {van Dokkum} (2007)
used spatially-resolved photometric and
dynamical observations of cluster early-type galaxies at $z\approx 0.5$
to directly test the assumption that the FP tracks $M/L$ evolution.
$M/L$ ratios derived from modeling
the spatially-resolved data
were directly compared to those determined from the FP.
Interestingly, the validity of the FP for determining
$M/L$ ratios appears to depend on the mass: for galaxies with
masses $\gtrsim 10^{11}$\,\msun\ or velocity dispersions $\gtrsim 200$\,\kms\
the $M/L$ ratios are consistent with each other, but for low mass
galaxies the FP appears to systematically underestimate the $M/L$ ratio.
Although these results suggest that
the traditional interpretation of FP evolution
is substantially correct at the high mass end, further
tests would be valuable, particularly at $z>0.5$.

Another complication is that the measured $M/L$ ratio
is not identical to the stellar $M/L$ ratio calculated in
stellar population synthesis models. In the models $M_*$ is
the sum of the mass of all living stars and the mass
of stellar remnants (black holes, neutron stars, and white dwarfs).
By contrast, the measured mass includes all possible contributions,
including dark matter and gas. Although the amount of dark matter
in elliptical galaxies is still somewhat uncertain
(e.g., {Romanowsky} {et~al.} 2003; {Mamon} \& {{\L}okas} 2005; {Cappellari} {et~al.} 2006), its
contribution within the effective radius is probably small
({Mamon} \& {{\L}okas} 2005; {Koopmans} {et~al.} 2006). The contribution of gas to the
total mass is also uncertain. One would expect substantial
amounts of gas and dust from stellar
mass loss (see, e.g., {Goudfrooij} {et~al.} 1994), qualitatively
consistent with Spitzer observations ({Temi}, {Brighenti}, \& {Mathews} 2007). However,
the data appear to be inconsistent with simple expectations,
and {Temi} {et~al.} (2007) suggest that the gas is transported outward,
possibly due to AGN activity.  Furthermore, in galaxy clusters the galaxies
move through very hot, diffuse gas at great speed and
it seems likely that any cold gas coming from stellar winds
is fairly efficiently stripped or heated.

Although dark matter and gas could contribute to the measured
$M/L$ ratio, these contributions will to large extent cancel
when comparing galaxies over a range of redshifts. In
the {Maraston} (2005) models the amount of mass loss over
the relevant age range from 5\,Gyr to 12\,Gyr is only $\approx 2$\,\%
for a {Salpeter} (1955) IMF and $\approx
3$\,\% for a {Kroupa} (2001) IMF. Even if
half of this mass was retained rather than lost to the intracluster
medium, the $M/L$ evolution would change by only $1-2$\,\%.

\subsection{Uncertainties in Stellar Population
Synthesis Models}

Uncertainties in the stellar population synthesis models are perhaps
the largest source of error, and at the same time they are notoriously
difficult to quantify. The fundamental problem is that the main
direct tests and calibrations of the models are offered by
Galactic open clusters and globular clusters, but that these
may not be representative for the integrated stellar
populations of massive early-type galaxies.

Recently the confidence in these
models was somewhat shaken by the large differences that exist
between
predictions of {Bruzual} \& {Charlot} (2003) and {Maraston} (2005). As discussed in
detail by {Maraston} (2005) and
others (e.g., {van der Wel} {et~al.} 2006b; {Maraston} {et~al.} 2006),
the predicted
rest-frame optical to near-infrared colors of simple stellar
populations can deviate by $\gtrsim 0.5$ magnitudes for identical
input metallicities and ages. The differences are in large part due
to a different numerical implementation of the thermally pulsating asymptotic
giant branch (TP-AGB) phase of stellar evolution, and are therefore
largest for ages 0.5--2 Gyr and colors which include a near-infrared band
(such as $B-K$).\footnote{It should be noted that
the treatment of TP-AGB stars is revised
in the most recent incarnation of the Bruzual \& Charlot
models, which brings them
more in line with the {Maraston} (2005) models
(see {Bruzual} 2007).}

\vbox{
\begin{center}
\leavevmode
\hbox{%
\epsfxsize=8.5cm
\epsffile{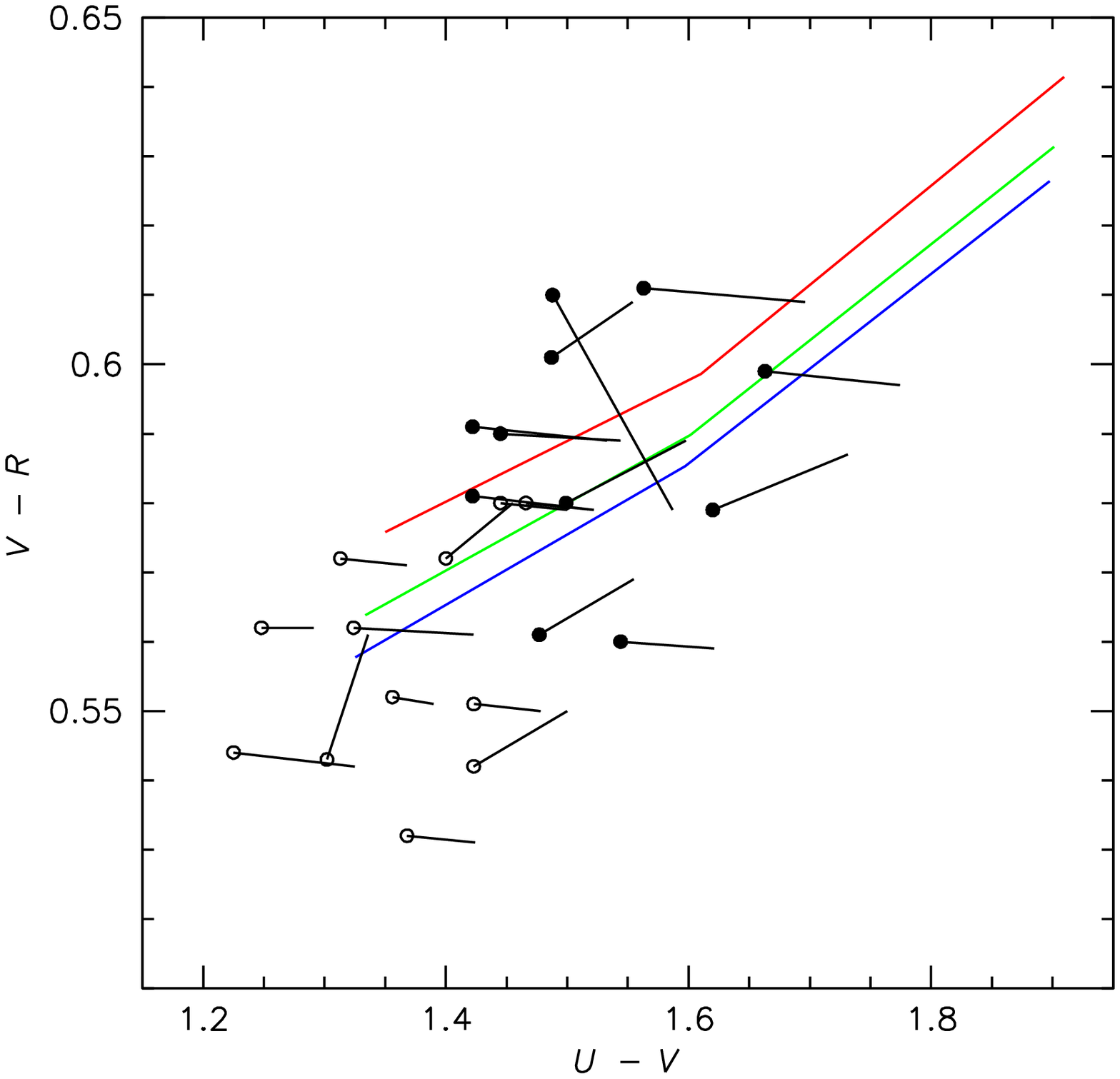}}
\figcaption{\small
Color-color diagram for galaxies in the Coma cluster. Colors were
obtained from Eisenhardt et al.\ (2007), for galaxies in common
with the J\o{}rgensen et al.\ (1995a,b) FP sample. Solid symbols
indicate galaxies with masses $>10^{11}$\,\msun. The lines attached
to the points show
the effects of changing the photometry aperture from $10\farcs 3$ to
$6\farcs 2$. Colored lines show model predictions from Maraston (2005)
for a 12 Gyr model with metallicity $-0.33\leq {\rm [Fe/H]} \leq +0.35$.
Different colors indicate different IMF slopes: $x=1.3$ (red),
$x=0.7$ (green), and $x=0$ (blue). The models fit the observed
colors well.
\label{colcol.plot}}
\end{center}}

Our analysis avoids the region of parameters space where the differences
between the models are most pronounced (ages of 0.5--2\,Gyr and 
rest-frame near-infrared passbands). The models are better
calibrated (and in much better agreement) for ages $>3$\,Gyr and rest-frame
optical passbands, largely because the contribution
of giants to the integrated light is significantly smaller.
However, subtle differences between the models also exist
at ages $>3$\,Gyr; as an example,
the $B-K$ colors of the {Maraston} (2005) models are slightly
bluer than those of {Bruzual} \& {Charlot} (2003) (see {van der Wel} {et~al.} 2006b).

To get some handle on the effects of the choice of
model the derivation of Eq.\ \ref{kappa3.eq} was
repeated for a Solar metallicity {Bruzual} \& {Charlot} (2003) model.
By comparing the color evolution to the $M/L$ evolution for
two different IMF slopes ($x=1.35$ and $x=0.35$) we find
$\kappa_B/\kappa_{U-V} \approx 5.0 -1.2 x$.
The relation is
less steep than that of the Solar metallicity {Maraston} (2005) model,
and is close to Maraston's super-Solar metallicity model (the solid
line in Fig.\ \ref{direct.plot}). For a Salpeter IMF and
Solar metallicity the {Bruzual} \& {Charlot} (2003)
model provides an even worse fit to the data than the {Maraston} (2005)
model, and the measured slope of the $M/L$ -- color relation
implies $x\sim -0.4$ (compared to $x\sim -0.1$ for the Solar
metallicity Maraston model). We infer that our conclusions are robust in the
context of presently popular models, but note that there are
significant variations in the model predictions even in the rest-frame
optical.

Finally, an important check on the entire enterprise
is whether the models can match the absolute colors of the galaxies.
So far, we have not used the absolute colors in any way and only
considered the rate of color {\em evolution}, leaving the normalization
a free parameter.
The absolute colors are difficult to interpret, as they suffer from
strong degeneracies between age, metallicity, and dust content.
Nevertheless, the models should be able to
reproduce the observed colors of
massive galaxies in Coma for plausible
combinations of these parameters. Figure \ref{colcol.plot}
shows the location of galaxies in the Coma cluster in a
color-color diagram of $V-R$ vs.\ $U-V$, along with
model predictions of {Maraston} (2005). All models
have an age of 12\,Gyr, corresponding to $z_{\rm form} \approx 4$.
The models have approximately the correct $U-V$ colors at
$z=0$ for $z_{\rm form}\approx 4$,
and are therefore self-consistent.
The fact that the agreement is particularly good for 
models with Solar metallicity and
$x=0$ is probably coincidence, given the uncertainties.

\subsection{Blue Stars in Old Stellar Populations}

As is well known old stellar populations can include very luminous
hot stars. These stars are rare, but due to their high luminosity they
can have
a significant effect on the integrated luminosity of a stellar
population, particularly in blue passbands.
These hot stars fall in three broad categories: blue horizontal branch
stars (e.g., {Zinn}, {Newell}, \& {Gibson} 1972; {Rich} {et~al.} 1997), extreme horizontal branch
stars (e.g., {Dorman}, {Rood}, \& {O'Connell} 1993), and blue stragglers (BSs)
(e.g., {Sandage} 1953; {Bailyn} 1995). Blue horizontal branch stars are
generally included in
current stellar population synthesis models.
Extreme horizontal branch stars are thought to be responsible
for the UV-upturn in elliptical galaxies at $\lambda\lesssim 2000$\,\AA\
({Burstein} {et~al.} 1988),
and should not affect the $U-V$ colors
(see, e.g., {Yi} {et~al.} 2003, 2005).

Blue stragglers
are not included in stellar population synthesis models,
probably because conditions for their formation
are still not well understood.
However, they can have a significant effect on the integrated near-UV
and optical colors of a stellar population. 
{Xin} \& {Deng} (2005) find that the contribution of BSs
to the integrated $B-V$ color of old open clusters
is typically $\sim 0.2$ magnitudes. Their contribution to the
$U-V$ color has not been measured directly, but is expected to
be $\sim 0.3$ magnitudes or so. In practice,  BSs
are not included when fitting isochrones to the stellar
populations of open clusters or globular clusters, but we do not
have that luxury for unresolved stellar populations.

The effect  of a BS
population such as that in M67
would be larger than the IMF
effects discussed in this paper, and
there is no a priori
reason to suppose that BSs do
not exist in elliptical galaxies (see, e.g., {Stryker} 1993).
Nevertheless, they probably do not
have a very large effect on our results. First, the relevant
question is not how much BSs contribute to the integrated
colors, but to what extent they influence their {\em evolution}
from $\sim 5$\,Gyr (the age at $z=0.83$ for
$z_{\rm form} \sim 4$) to $\sim 12$\,Gyr (the age at $z=0$).
To have a significant impact on our results
BSs would have to have been
absent in elliptical galaxies for five or more Gyr and then
have ``turned on''. This seems rather contrived,
particularly since the turn-off mass changes by
only $\sim 0.4$\,\msun\ over this age range. If anything,
one might expect BSs to be slightly more prevalent at younger
ages. One of the preferred models for their formation\footnote{At
least in old open clusters; other mechanisms may dominate at young
ages and in other environments.}
is through the coalescence of
binary stars ({Mateo} {et~al.} 1990), and if binaries
start out with a flat distribution of orbital parameters the
rate at which some of them turn into BSs will gradually decrease. 
Coupled with the fact that BS lifetimes are short
compared to the main sequence stars, this could lead to
a gradual decline in the blue straggler population.
More to the point,
{Xin} \& {Deng} (2005) have determined the contribution of BSs
to the integrated light of open clusters with a large range
of ages. They find no correlation between the contribution of
BSs to the integrated colors
and the age of the cluster. As a result, the integrated colors
of clusters with ages varying from 1\,Gyr to
8\,Gyr are reasonably well described
by a single stellar population synthesis
model with a constant offset.

Another reason why the effects of BSs are probably fairly mild
is that it is difficult to hide them completely in nearby
elliptical galaxies. If BSs are as prevalent in elliptical galaxies as
they are in open clusters, they would lead to relatively
blue near-UV colors and strong Balmer absorption lines.
Population synthesis models
would (correctly) infer the presence of
A and F main sequence stars, and (incorrectly) attribute them
to a secondary young stellar population, and/or assign an overall
young luminosity-weighted age to the galaxy.
As pointed out by {Schiavon}, {Caldwell}, \&  {Rose} (2004), 
if a galaxy had the same stellar population as M67 it would be
classified as ``E+A'' ({Dressler} \& {Gunn} 1983)
to signify the presence of a post-starburst
component in addition to an older component.
{Xin} \& {Deng} (2005) quantify these effects and conclude that the ages
of unresolved stellar populations will typically be underestimated by a
factor of $\sim 2$ if blue stragglers are present in similar numbers
as in open clusters.
Turning to many years of population synthesis modeling of massive
elliptical galaxies in clusters, such dramatic effects can be
safely ruled out. As an example,
{Thomas} {et~al.} (2005) infer ages of $\sim 12$\,Gyr (i.e., close
to the age of the Universe)
for cluster galaxies with $M\gtrsim 10^{11}$\,\msun, not
$\sim 6$\,Gyr as one might expect from a BS-rich stellar
population. 
Qualitatively,  the red colors and weak Balmer lines of
massive cluster ellipticals place strong limits on
the relevance of BSs in these galaxies.

Finally, we note that {Trager} {et~al.} (2000b) argue that BSs do not contribute
significantly to the integrated light
even in elliptical galaxies that {\em do} have enhanced
Balmer lines and signs of young populations (typically relatively
low mass field ellipticals).
Rather than discuss their arguments, we note that the finding that
the derived age of ellipticals
appears to correlate with
the presence of morphological fine structure ({Schweizer} \& {Seitzer} 1992)
provides additional support for the notion that late star formation,
rather than a BS population, is generally responsible for the young
appearance of some nearby ellipticals.
We are left to wonder {\em why} BSs seem to be deficient in
elliptical galaxies, as compared to old open clusters.
No speculation will be offered here, as this question is
well outside the scope of the present study.

\subsection{Complex Evolution and Dust}

As argued in \S\,\ref{implem.sec} the population of
massive early-type galaxies in clusters
is not thought to have experienced significant star
formation at redshifts $\ll 1$. Young stellar populations
have been seen in some field ellipticals
(e.g., {Trager} {et~al.} 2000a; {Thomas} {et~al.} 2005), but a large body of observational
evidence supports the notion that massive cluster galaxies are
remarkably homogeneous (e.g., {Bower} {et~al.} 1992a; {Stanford} {et~al.} 1995, 1998; {Ellis} {et~al.} 1997; {Kodama} {et~al.} 1998; {van Dokkum} {et~al.} 2000; {Holden} {et~al.} 2004; {Thomas} {et~al.} 2005; {Tran} {et~al.} 2007).
This apparent homogeneity may be misleading
and mask more complex evolution
({van Dokkum} \& {Franx} 2001), but this is not likely
for the most massive galaxies ({Holden} {et~al.} 2007).
It may well be that cluster early-types experienced more
complex evolution at $z>1$, in particular during and before
virialisation of the clusters in which they now live, but this
does not affect the analysis. The parameter
$z_{\rm form}$ does not refer to a single star formation event,
but reflects the luminosity-weighted mean formation epoch of
the stars. As shown explicitly in {van Dokkum} {et~al.} (1998b) color
and luminosity evolution are nearly identical for
stellar populations with different
star formation histories but the same luminosity-weighted age,
as long as star formation terminated at least $\sim 1$\,Gyr
prior to the epoch of observation.

Dust needs to be treated separately, as significant amounts of
gas and dust are expected to exist in early-type galaxies as
a result of
winds emanating from massive stars
(see also \S\,\ref{whatml.sec}).
Cluster galaxies at $z\sim 0.8$ may still have retained a significant
fraction of the dust expelled by massive
stars, whereas it may have been stripped, blown out, or obliterated
at later times (see, e.g., {Goudfrooij} {et~al.} 1994; {Temi} {et~al.} 2007).
Quantifying these effects is very difficult, as they not only
depend on the amount of mass loss but also on the detailed processes and
timescales for cooling and heating of the dust. Qualitatively,
one might expect
higher reddening at $z\sim 0.8$ and therefore a lower
apparent color evolution from $z\sim 0.8$ to $z\sim 0$.
The true color evolution would then be stronger, and the data
might be consistent with a normal IMF in combination
with a relatively low stellar age.

However, dust would not only affect the colors but also the luminosities.
The relation between reddening and extinction in elliptical galaxies
is not well known, but is probably not very different
from that in the Milky Way (e.g., {Goudfrooij} {et~al.} 1994).
For reference, arrows in Fig.\ \ref{direct.plot} show the effects of
Milky Way-like dust and of the (relatively grey) {Calzetti} (1997)
extinction curve. Calzetti-like dust would have essentially no
effect on the analysis, as the dust vector is nearly parallel to
the observed relation between $M/L$ ratio and color.
Milky Way-like dust would have an effect, but would lead to
unrealistically low ages for the galaxies. To change the observed
relation to the one expected for a standard IMF the $z=0.8$
galaxies would have to be reddened by $\sim 0.3$ mag
in $U-V$, corresponding to a $B$-band extinction of $\sim 0.6$ mag.
The implied luminosity evolution from $z=0.8$ to the present would
then be $\sim 1.7$ magnitudes (instead of 1.1
magnitudes; see vv07). This
in turn would mean that massive cluster
galaxies formed their stars at 
$z\sim 1.3$, which is ruled out by many independent
observations (e.g., {Kelson} {et~al.} 2001; {Thomas} {et~al.} 2005; {Kodama} {et~al.} 2007).

\subsection{Summary of Fitting Results and Associated Uncertainties}
\label{fitsum.sec}

The main result from the fits in \S\,\ref{apply.sec} is that
stellar population synthesis models with
a standard, Salpeter-like IMF in the region around 1\,\msun\ 
are not able to simultaneously fit the luminosity and color evolution of
massive cluster galaxies.
By contrast,
models with  flat
IMFs and high stellar formation redshifts
provide excellent fits. 
The slow color evolution measured here is not
easily explained by invoking sample selection effects,
photometric errors,
a blue straggler population, or differential dust extinction.

The formal limits on the IMF slope and formation redshift
are obviously very uncertain, as they 
do not take any of the effects
discussed above into account.
Furthermore, negative IMF slopes are outside of the
parameter space covered by the models, which makes the
formal Super-solar metallicity solutions somewhat suspect.
In what follows
the formal Solar metallicity solutions
($x = -0.3^{+0.4}_{-0.7}$ and $z_{\rm form}=3.7^{+2.3}_{-0.8}$)
will be used as a ``default'' model, but it should be kept in
mind that the errorbars almost certainly underestimate the
true uncertainty.

\section{Comparison to Previous Results}
\label{comp.sec}

Many previous studies have determined or constrained the
properties of the stellar populations of early-type galaxies.
Although direct comparisons are difficult
because of differences in sample selection and methodology, it is
important to verify that our results are broadly consistent with
the large body of literature that exists on this topic.

\subsection{Other studies of the Evolution of $M/L$ Ratios and Colors}

As discussed in detail in vv07, the values shown in Fig.\
\ref{fitboth.plot}a are mostly determined from literature data, and
therefore ``automatically'' consistent with a large body of
previous work. The only exception is {J\o{}rgensen} {et~al.} (2006), who find
substantially slower evolution in the first version of their
published paper. However, this was due to a conversion error,
and the vv07 results are fully consistent with the Erratum
to {J\o{}rgensen} {et~al.} (2006) published in 2007.

The evolution of the rest-frame $U-V$ color cannot be directly
compared to previous work, as most studies choose
to compare the observed
colors to redshifted model predictions rather than
convert the data to the rest-frame. Furthermore, color
evolution has not been studied previously for mass-limited
samples of cluster galaxies. Nevertheless, it is interesting
to compare
the constraints on the stellar ages that are reported in
the literature, as they should reflect the measured rate of color evolution.

Most studies of the color evolution of early-type galaxies in
clusters find very old ages, in qualitative agreement with
the results found here (e.g., {Stanford} {et~al.} 1995, 1998; {Kodama} {et~al.} 1998; {Koo} {et~al.} 2005). In particular,
no studies have reported formation redshifts as low as
$\sim 2$  for the most massive galaxies in clusters, which is the
value implied by the evolution of the $M/L$ ratios for a standard
IMF. The most comprehensive study to
date is by {Holden} {et~al.} (2004), who combined data
from {Stanford} {et~al.} (1998) with new measurements to obtain a combined
sample of 31 clusters at $0.3<z<1.3$ with multi-band
photometry. They find
$z_{\rm form}=5^{+\infty}_{-2.7}$ from Solar metallicity {Bruzual} \& {Charlot} (2003)
models, in good agreement with our results.
It should be noted that the {Holden} {et~al.} (2004) study may suffer
from some of the same systematic effects as the work described
here; for example, an emerging blue straggler population would
influence all studies of color evolution.

\subsection{Evolution of Balmer Absorption Lines}

{Kelson} {et~al.} (2001) study the evolution of the Balmer absorption
line strengths of early-type galaxies in clusters at $0.06\leq z \leq 0.83$,
and their data offer an independent check on the
results derived here. The Balmer lines should behave like the
$U-V$ colors, in the sense that they are fairly insensitive to the
IMF and their rate of evolution is mostly determined by the age of the
stellar population. Importantly, the sources
of systematic error (both in the data and in the models) are
very different from those that affect the colors
(see, e.g., {Kelson} {et~al.} 2006).

The {Kelson} {et~al.} (2001) data are
shown in Fig.\ \ref{kelson.plot}. The data are compared to two
Solar metallicity {Bruzual} \& {Charlot} (2003) models: 
a model with a standard IMF
and a young age,  and a model with a top-heavy IMF and an old age.
As discussed in vv07 and in \S\,\ref{joint.sec} these are the only two classes
of models which provide good fits
to the observed evolution of the $M/L$ ratio.
The young model is clearly inconsistent with the data, and can be
rejected with high confidence. By contrast,
the top-heavy/old model fits the data very well.
This model is the formal best fit to the $M/L$ and color
evolution (as discussed in
\S\,\ref{fitsum.sec}), but other models with formation redshifts
in the range 3--6 also fit well.

The evolution of the Balmer lines thus
provides strong independent support for the analysis in
\S\,\ref{apply.sec}.  We note that
the inferred ages are consistent with the analysis
in {Kelson} {et~al.} (2001): they 
give 95\,\% confidence lower limits to $z_{\rm form}$
of 2.4 and 2.9 for two different stellar
population synthesis models.

\vbox{
\begin{center}
\leavevmode
\hbox{%
\epsfxsize=8.5cm
\epsffile{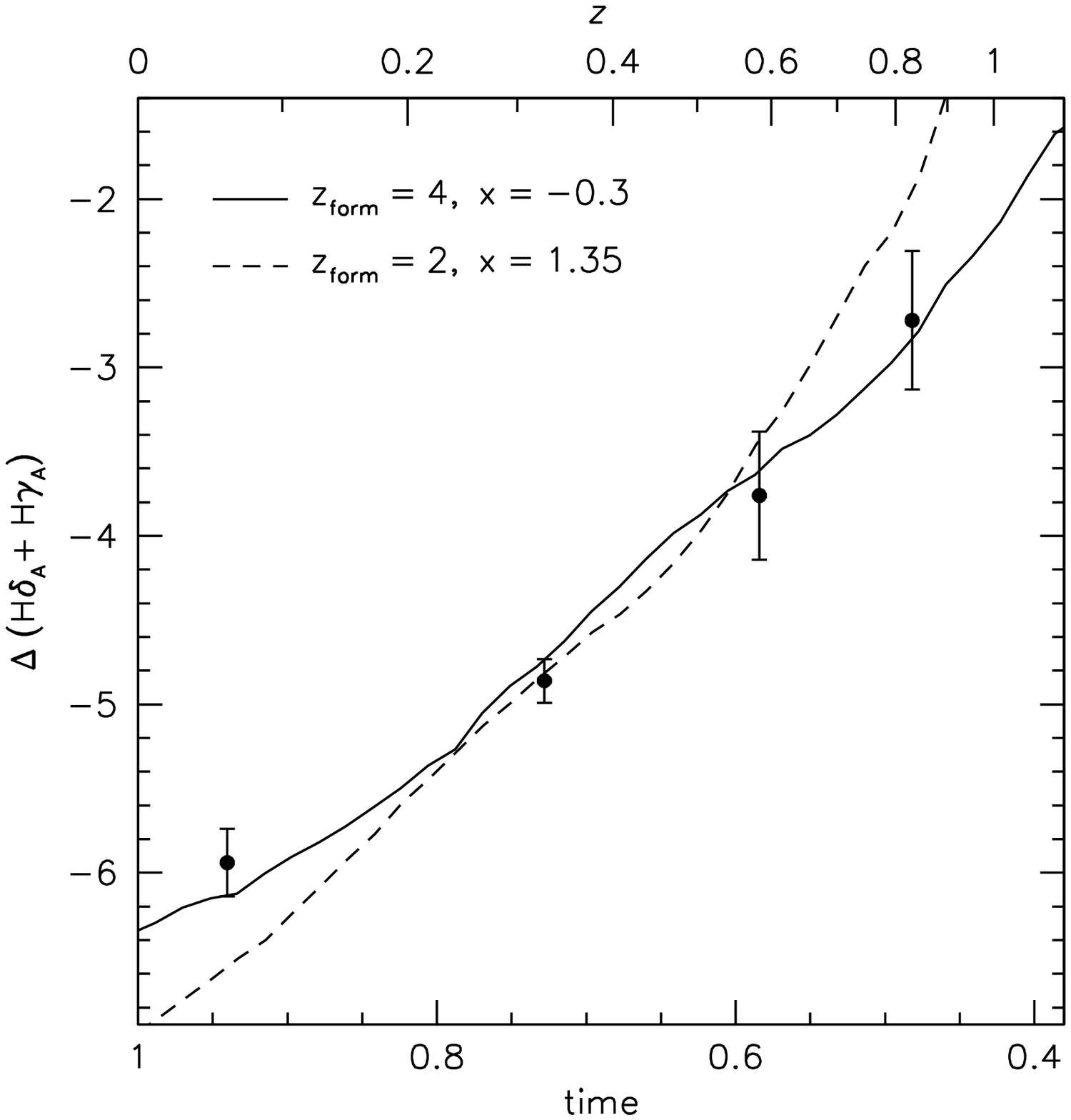}}
\figcaption{\small
Evolution of the Balmer absorption line strengths of early-type
galaxies in clusters, taken from Kelson et al.\ (2001). The
lines show the two classes of models that fit the evolution
of the $M/L$ ratio of early-type galaxies: young models
with a Salpeter IMF and old models with a top-heavy IMF.
Only the old/top-heavy model fits the
linestrength data, providing independent support
for our results.
\label{kelson.plot}}
\end{center}}

\subsection{Evolution of Field Early-Type Galaxies}

Field early-type galaxies probably evolve in more complex ways
than their counterparts in clusters.
Galaxy-galaxy
mergers are expected to be rare in virialized clusters
(e.g., {Makino} \& {Hut} 1997) but could be common in groups
(e.g., {van Dokkum} 2005; {Bell} {et~al.} 2006), and the same is
probably true for late star formation (e.g., {Trager} {et~al.} 2000a).
Field samples may therefore not be suitable for measuring
the subtle effects discussed in this paper. Nevertheless,
it is interesting to note that {van der Wel} {et~al.} (2006a) find
that {Bruzual} \& {Charlot} (2003) models with a standard IMF do not fit
the observed relation between $B-K$ color and $M/L_B$ ratio
of field early-type galaxies at $0<z<1$. Models with top-heavy
IMFs fit better, although {Maraston} (2005) models with a
standard IMF also provide good fits.

\subsection{``Red and Dead'' Galaxies at High Redshift}

Recent studies of faint $K-$selected samples have identified a population
of apparently ``red and dead'' galaxies at
$z\approx 2.5$ ({Franx} {et~al.} 2003; {Labb{\'e}} {et~al.} 2005; {Daddi} {et~al.} 2005; {Kriek} {et~al.} 2006). Deep rest-frame
optical spectra demonstrate strong Balmer breaks and no line emission
in a substantial fraction of massive galaxies at these redshifts,
implying that star formation in these objects
mostly terminated at $z\approx 3$ or
earlier ({Kriek} {et~al.} 2006). These objects are found
in blind surveys of relatively small fields, and it seems likely
that many galaxies in (proto-)clusters terminated star formation
even earlier, and/or that the fraction of (nearly) passive galaxies
is higher in those environments (see, e.g., {Thomas} {et~al.} 2005; {Steidel} {et~al.} 2005; {Quadri} {et~al.} 2007; {Kodama} {et~al.} 2007). Assuming that
these galaxies are not ``rejuvenated'' at later
times (which is entirely possible),
the presence of substantial numbers of apparently
old galaxies at this redshift places an interesting constraint
on the luminosity-weighted ages of massive
early-type galaxies.

For any star formation history the time of {\em last}
star formation will be more recent than the
luminosity-weighted mean star formation epoch. After
a $\delta$-function, the
simplest star formation history
is a top-hat with a constant star formation rate from $t=t_{\rm start}$
to $t=t_{\rm stop}$. Such models may actually be fairly realistic
(as compared to exponentially declining models, for instance), as they
superficially resemble the star
formation histories of massive galaxies in hydrodynamical simulations
({Nagamine} {et~al.} 2005). As shown in {van Dokkum} {et~al.} (1998b), there is a simple
relation between the luminosity-weighted age $t-t_{\rm form}$
and $t_{\rm stop}$ in these models: $t-t_{\rm form} \approx
\sqrt{(t-t_{\rm start})(t-t_{\rm stop})}$. Assuming that
$z_{\rm stop}\approx 3$ (extrapolating from the Kriek et al.\ results
for red field galaxies), the luminosity-weighted mean formation
epoch $z_{\rm form}\gtrsim 3.7$ for 
$t_{\rm stop}-t_{\rm start}\gtrsim 1$\,Gyr.

Luminosity-weighted
formation redshifts of $\sim 4$ are entirely consistent with the
old/top-heavy model that is implied by the analysis in \S\,\ref{apply.sec}.
Formation redshifts of 2--2.5, which are implied by the $M/L$
ratio evolution for Salpeter-like IMFs (see vv07 and
\S\,\ref{apply.sec}) may be difficult to reconcile with the presence
of apparently old field galaxies at $z\sim 2.5$.
We note, however,
that this type of evidence should be treated with caution,
as the old high redshift objects could constitute only a small fraction of
the progenitors of today's early-type galaxies.

\subsection{Linestrengths of Nearby Elliptical Galaxies}

Many studies have constrained the properties of the stellar populations
of nearby early-type galaxies using absorption line strengths
(e.g., {Peletier} 1989; {Worthey} {et~al.} 1992; {Trager} {et~al.} 2000a; {Thomas} {et~al.} 2005).
Linestrengths
are generally more helpful than colors, as they suffer slightly
less from the strong degeneracies that exist between
(particularly) age and metallicity (e.g., {Worthey} 1994).
One of the difficulties in linestrength studies is that
the abundance ratios of early-type galaxies do not match Solar
models ({Peletier} 1989),
which in itself may be indirect evidence for a top-heavy
IMF at early times ({Worthey} {et~al.} 1992). {Thomas} {et~al.} (2005)
argue that models which explicitly include non-Solar abundance ratios
can date the star formation epoch of early-type galaxies. These
models indicate
that cluster early-type galaxies formed their stars at $z=3-5$,
in good agreement with the ages derived here for top-heavy IMFs.

\subsection{Absolute M/L Ratios of Nearby Elliptical Galaxies}
\label{absml.sec}

In \S\,\ref{apply.sec} we were only concerned with the evolution
of the derived $M/L$ ratios,
and not with their absolute values. The
absolute $M/L$ ratios of nearby elliptical galaxies provide
a powerful additional constraint on the
IMF, as the stellar $M/L$ ratio implied by the combination of
the IMF and the observed
luminosity should not exceed the dynamical $M/L$ ratio.
As discussed in \S\,1 the observations of {Cappellari} {et~al.} (2006)
are consistent with the stellar
mass implied by a {Chabrier} (2003) IMF and rule out IMFs
which are significantly steeper than the Salpeter value, because
they have too much mass locked up in low mass stars.

Perhaps surprisingly, the dynamical $M/L$ ratios also
rule out top-heavy IMFs of the form fitted in \S\,\ref{apply.sec}.
These IMFs have less low mass stars than a standard
{Salpeter} (1955) IMF but many more high mass stars.
As a result, the stellar mass function at
late times is ``remnant heavy'', dominated as it is by
white dwarfs, neutron stars, and black holes. {Maraston} (1998)
shows that the $M/L$ ratio at old ages reaches a minimum
for $x\sim 1.3$: for larger values the IMF is dwarf-dominated
and for smaller values the IMF is remnant-dominated.
Specifically, Solar metallicity
{Maraston} (2005) models predict
$M/L_B = 9$ at 12\,Gyr for $x=1.35$ and $M/L_B = 87$ for
$x=0$! Such high $M/L$ ratios are an order of magnitude
higher than measured
(see {van der Marel} 1991; {Cappellari} {et~al.} 2006; {Gavazzi} {et~al.} 2007, and many other studies), which
means that IMFs with a constant logarithmic slope $x=0$
over the entire mass range $0.1-100$\,\msun\ can effectively be
ruled out. In the next Section we will explore IMFs which
have a Salpeter-like slope at high masses and
$x\sim 0$ in the region around 1\,\msun. Such IMFs
are physically more plausible than IMFs with a constant
slope and do not violate dynamical limits on the
stellar $M/L$ ratios of nearby galaxies (see
\S\,\ref{mass.sec}).

\section{Implications}
\label{imply.sec}
\subsection{Implications for the Characteristic Mass at $z \sim 4$}

\label{chabrier.sec}

In the {Maraston} (2005)
models described in \S\,\ref{starpop.sec} the IMF is parameterized
as a power law with constant slope $x$
over the entire mass range $0.1 - 100$\,\msun.
The best-fitting slope of $x\sim 0$ implies an extremely top-heavy
mass function, with a very large number of massive stars.
However, other forms of the IMF are also consistent with the data,
as the analysis of \S\,\ref{apply.sec}
is completely insensitive to the IMF at masses exceeding the turn-off
mass at $\sim 5$\,Gyr ($\approx 1.15$\,\msun). A
physically more plausible IMF would have a {Chabrier} (2003)
form, with a fixed high-mass slope of $x\approx 1.3$
and a varying characteristic mass $m_c$.
As will be quantified later, such IMFs are not only physically
motivated but are also entirely consistent with
the dynamical $M/L$ ratios measured for nearby ellipticals.

So far, no stellar population
synthesis modeling has been done with Chabrier
IMFs with varying $m_c$.
However, we can use the fact that our observations probe the IMF
only over a limited stellar mass range to
derive an approximate relation between $m_c$
and the slope of the IMF as parameterized in the {Maraston} (2005)
models.
The {Chabrier} (2003)
IMF has the form
\begin{equation}
\label{chabrier.eq}
\xi = \left\{
\begin{array}{c@{\quad}l}
A_l \exp\left[-(\log m - \log m_c)^2/2\sigma^2\right] & (m\leq 1\,M_{\odot}) \\
A_h m^{-x} & (m>1\,M_{\odot}),
\end{array} \right.
\end{equation}
with $A_l = 0.158$, $m_c = 0.079$\,\msun, $\sigma = 0.69$,
$A_h = 0.0443$,
and $x=1.3$, and is shown by the dashed lines in Fig.\
\ref{chabrier.plot}. As shown in {Chabrier} (2003) this IMF is
very similar to the {Kroupa} (2001) IMF, and both IMFs are
nearly identical to the {Salpeter} (1955) IMF for $m>1$\,\msun.
The bottom panel of Fig.\ \ref{chabrier.plot}
shows the derivative $d(\xi
(\log m))/d(\log m)$, which can be thought of as the local
logarithmic slope of the IMF. 
The slope gradually
steepens until leveling off at the Salpeter value for masses
$m>1$\,\msun.

\vbox{
\begin{center}
\leavevmode
\hbox{%
\epsfxsize=8.5cm
\epsffile{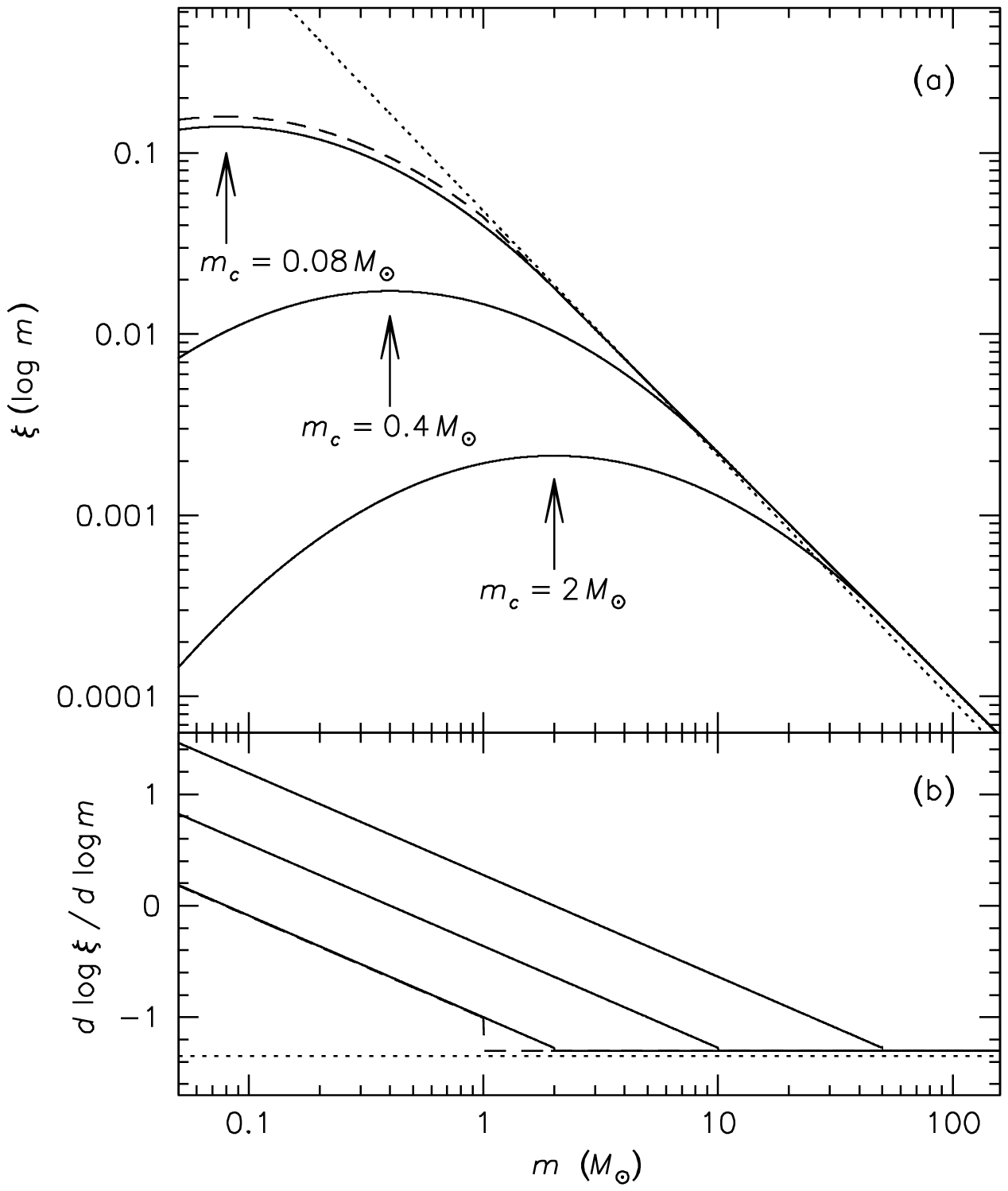}}
\figcaption{\small
Different forms of the IMF (a), and their derivatives (b).
Dotted lines indicate the Salpeter (1955) IMF, with constant logarithmic
slope $x=1.35$. The dashed lines show the Chabrier (2003)
disk IMF, which turns over at $m_c \approx 0.08$\,\msun.
Solid lines show a slightly modified form of the Chabrier (2003)
IMF, which eliminates the discontinuity at $m=1$\,\msun\
and explicitly accommodates a varying $m_c$.
\label{chabrier.plot}}
\end{center}}

The derivative of
the {Chabrier} (2003) IMF has a discontinuity at
1\,\msun, where the log-normal part of the IMF and the power law
part connect. This discontinuity is rather minor and has no
practical implications, but it
becomes more pronounced when considering higher values of $m_c$.
Therefore, we adopt an extension of the
{Chabrier} (2003) disk IMF parameterization
which does not produce
discontinuities in the derivative and explicitly accommodates
a varying $m_c$:
\begin{equation}
\label{mod.eq}
\xi = \left\{
\begin{array}{c@{\quad}l}
A_l (0.5 n_c m_c)^{-x}
\exp\left[-(\log m - \log m_c)^2/2\sigma^2\right] & (m\leq n_c m_c) \\
A_h m^{-x} & (m>n_c m_c),
\end{array} \right.
\end{equation}
with $A_l = 0.140$, $n_c = 25$, and $\sigma$, $A_h$, and $x$ identical
to Eq.\ \ref{chabrier.eq}
(see {Chabrier} (2003) for an alternative parameterization
of IMFs with a high characteristic mass). The behavior
of this modified Chabrier IMF for $m_c=0.08$\,\msun,
$m_c=0.4$\,\msun, and $m_c=2$\,\msun\
is shown in Fig.\ \ref{chabrier.plot}(a) and (b).
For $m_c=0.08$ this IMF is nearly identical to the
{Chabrier} (2003) disk IMF.

The observations discussed in this paper are sensitive to the IMF
in a limited stellar mass range near 1\,\msun.
The $B$-band luminosity-weighted mean stellar mass of a stellar
population is typically slightly smaller than the turn-off mass, and
ranges from 1.0--0.8\,\msun\ for ages of 5--12\,Gyr.
As is evident from
Fig.\ \ref{chabrier.plot} the IMF can be approximated by a powerlaw
over this small mass range, and the slope of this powerlaw
depends on $m_c$. We find
\begin{equation}
\label{mc.eq}
\log m_c \approx -0.05 -1.1 x_1 \quad \quad (x_1<1.3)
\end{equation}
with $x_1 = -d\log \xi / d \log m$ the logarithmic slope of the IMF near
$m \approx 1$\,\msun.
This expression relates 
Salpeter-like IMFs with varying
slope $x$, as used in the {Maraston} (2005) models and 
Eqs.\ \ref{kappa1.eq} and \ref{kappa2.eq},
to physically more plausible
Chabrier-like IMFs with fixed high mass slope
and varying characteristic mass.

The best-fitting slope of the IMF as derived from
Solar metallicity {Maraston} (2005) models is
$x=-0.3^{+0.4}_{-0.7}$
(see \S\,\ref{fitsum.sec}). In the context of a Chabrier-like
IMF this slope implies a characteristic mass
$m_c = 1.9^{+9.3}_{-1.2}$\,\msun\ at $z=3.7^{+2.3}_{-0.8}$. For
super-Solar metallicity the best-fitting values of $x$ are
strongly negative and may imply $m_c \sim 10$\,\msun, although
this represents a substantial extrapolation of the models.

We conclude that the color- and luminosity evolution of early-type
galaxies are consistent with IMFs which have a Salpeter slope at
high masses but turn over near $\sim 1$\,\msun. Such an IMF is
perhaps best described as ``bottom-light'' rather than top-heavy:
it does not have a larger number of massive stars than a standard
{Chabrier} (2003) IMF, but has a deficit of low mass stars.
The form of the IMF at masses $<1$\,\msun\ is not
constrained by the data presented here, and it is an open
question whether the slope of the
IMF actually becomes negative at low masses or stays constant
at $x\sim 0$.

\subsection{Evolution of the Characteristic Mass}

The evolution of the characteristic mass is shown in Fig.\
\ref{larson.plot}. The results obtained in this study are
represented by the formal best fit
for Solar metallicity, as discussed in
\S\,\ref{joint.sec}. The point labeled 'SMGs'
reflects results of {Blain} {et~al.} (1999a) for
submm galaxies (SMGs). 
{Blain} {et~al.} (1999a) find that a standard
{Salpeter} (1955) IMF has too many low mass stars,
which would produce too much $K$-band light at $z=0$
a similar argument was made by {Dwek} {et~al.} (1998, based on COBE data).
Blain et al.\  find that a top-heavy IMF with a
simple cutoff at 1\,\msun\ resolves this discrepancy.
No errorbar is given, but we can assume that
the uncertainty in the amount of
``missing mass'' is at least 50\,\%. 
The value we adopt for $m_c=0.34$, to match the
total stellar mass of the truncated
Salpeter (1955) IMF invoked by {Blain} {et~al.} (1999a).
The redshift comes from {Chapman} {et~al.} (2003).

The Figure also includes the IMFs of the Milky Way disk
(from {Chabrier} 2003) and
of globular clusters.
The age for globular clusters
reflects the results of {Gratton} {et~al.} (2003), who find that
globular clusters in the inner halo have ages ranging from
$\sim 13.4$\,Gyr to $\sim 10.8$\,Gyr with a 
systematic uncertainty of $\approx 0.6$\,Gyr.
The characteristic mass was determined by
{Paresce} \& {De Marchi} (2000),
who infer a typical characteristic mass $m_c = 0.33$,
with formal uncertainty $\pm 0.03$. The true uncertainty
is probably somewhat larger, as dynamical effects may play
a role even at low masses (see, e.g., {de Marchi},
{Paresce}, \& {Portegies  Zwart} 2005). The Galactic bulge is
not included; no estimates have been made for the characteristic mass
of the bulge IMF, but we note that there is evidence that the IMF
in the range 0.15 -- 1\,\msun\ 
is flatter than a Salpeter IMF (Zoccali et al.\ 2000).

The broken lines in Fig.\ \ref{larson.plot} are examples of models in which
the temperature of the CMB effectively sets the characteristic mass
scale of star formation at high redshift
(see {Larson} 1998, 2005; {Jappsen} {et~al.} 2005).
The CMB temperature exceeds the typical temperatures in prestellar
cores in molecular
clouds of $\sim 8$\,K (e.g., {Evans} {et~al.} 2001) beyond $z\sim 2$.
As discussed in {Larson} (2005) the exact temperature dependence
is very uncertain, and can vary from $\propto T^{1.7}$ ({Jappsen} {et~al.} 2005, dotted
line) to $\propto T^{3.35}$ ({Larson} 1985, dashed line).
The normalization of these models is also very uncertain. The value
of $\sim 0.3$\,\msun\ is the approximate value of the Jeans mass in present-day
cores ({Larson} 2005), but other aspects may play a role in setting
the characteristic mass at low redshift
(see {McKee} \& {Ostriker} 2007). The grey solid line is a non-physical
``toy model'' which fits the Milky Way disk, submm-galaxies,
and cluster galaxies and is intermediate between the two
physically-motivated models at high redshift. This model has
the form
\begin{equation}
\label{toy.eq}
m_c^2 = 0.08^2  \left(1+ \left( T_{\rm CMB}/6 \right)^6 \right)
\end{equation}
and is used to explore the effects of an evolving characteristic mass
in the following Sections.

\vbox{
\begin{center}
\leavevmode
\hbox{%
\epsfxsize=8.5cm
\epsffile{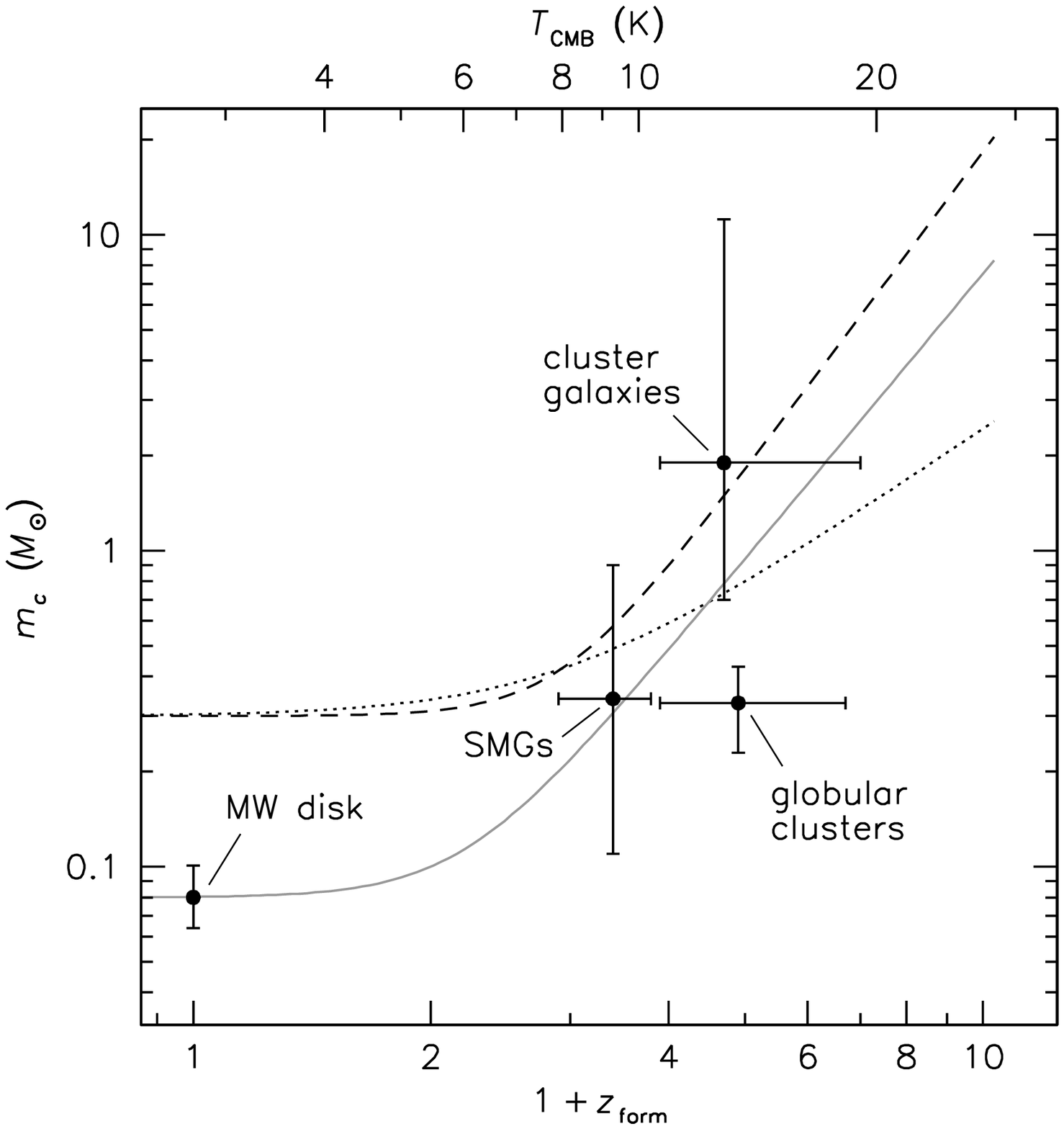}}
\figcaption{\small
Estimates of the characteristic mass of the IMF, as a function of
the epoch of star formation. The point marked ``cluster galaxies''
is the best-fit
Solar metallicity model to the observed evolution of massive early-type
galaxies in clusters. Also shown are
estimates for the present-day Milky
Way disk, for globular clusters in the inner halo, and
for submm-galaxies (SMGs).
Broken lines are examples of models of Larson (2005),
in which the characteristic mass 
is a strong function of the
temperature of the microwave background radiation
at suffuciently high redshift.
The dashed (dotted) line shows a model with
a $\propto T^{3.35}$ ($\propto T^{1.7}$) dependence at
$z>2$. The grey solid line is a non-physical model which fits the
available data.
\label{larson.plot}}
\end{center}}

It is interesting that the IMF of old globular clusters appears to have
a lower characteristic mass than the IMF of massive cluster galaxies,
even though their stars probably have similar mean ages.
Determining the IMF in globular clusters is notoriously difficult,
as the mass function as observed today is heavily
influenced by dynamical effects in combination with stellar evolution
(see, e.g., {Chabrier} 2003). In particular, globular clusters with
a very top-heavy IMF would not be expected to survive
in the tidal field of our Galaxy
for a Hubble time (e.g., {Joshi}, {Nave}, \& {Rasio} 2001), and the clusters that
are surviving today might not be representative for the original
population. Nevertheless,
taking the existing constraints at face value, it seems that other
parameters than the CMB temperature
(e.g., metallicity, star formation rate, or mass
of the star forming complex) may play a role in determining $m_c$
at high redshift.

\vbox{
\begin{center}
\leavevmode
\hbox{%
\epsfxsize=8.5cm
\epsffile{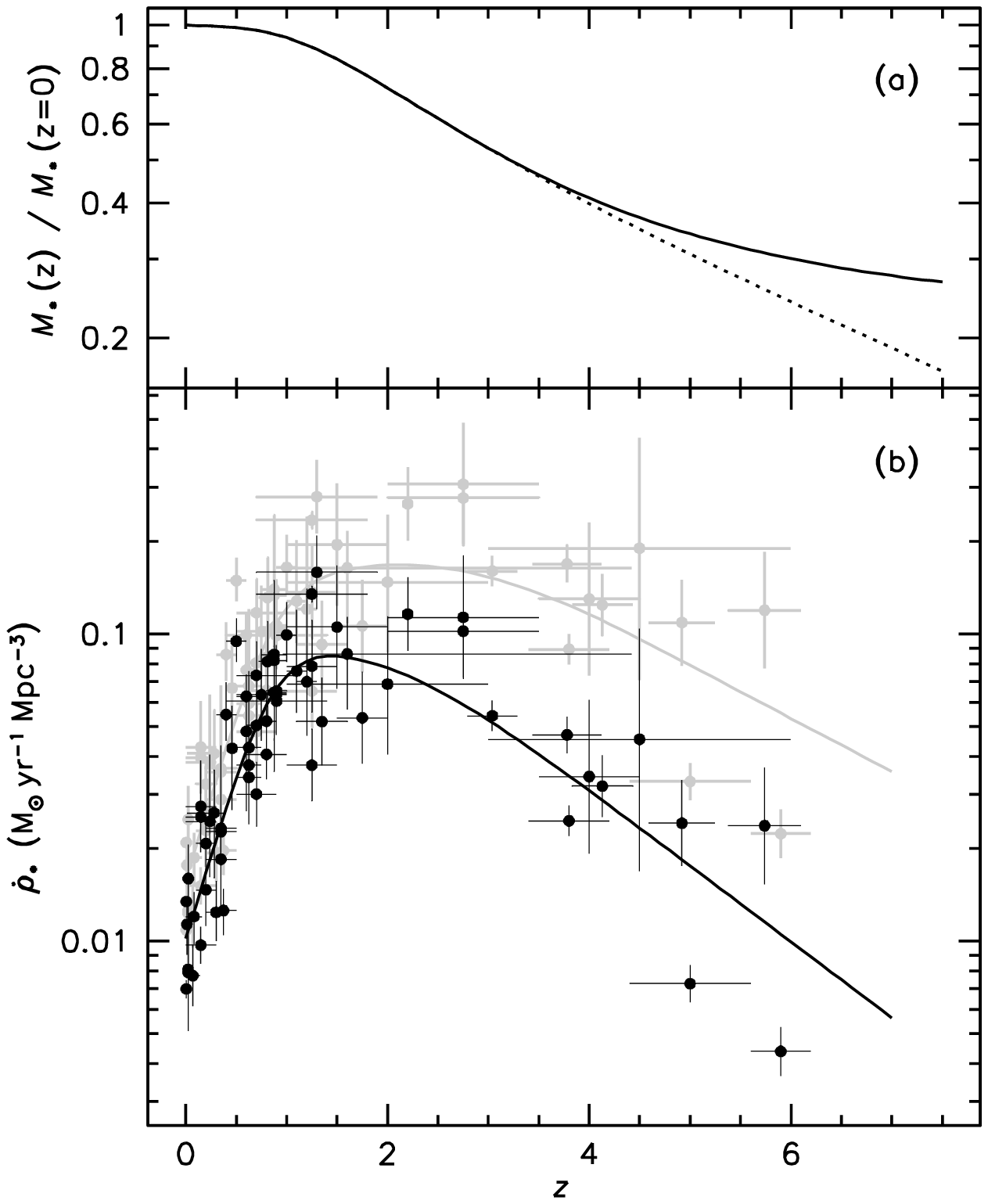}}
\figcaption{\small
Effects of an evolving IMF on the cosmic star formation history.
The dashed line in (a) shows the change in the total initial stellar mass
with redshift, if the characteristic mass $m_c$ evolves according
to Eq.\ \ref{toy.eq}. The solid line shows the total mass relative
to the mass in stars of $>10$\,\msun, which supply most of the UV
luminosity. (b) Measurements of the volume-averaged star formation
rate, from Hopkins (2004) and Bouwens et al.\ (2007). Grey points
are for a Salpeter (1955) IMF, and black points are for a
Chabrier (2003) IMF with evolving $m_c$. The modified evolution
shows a peak at $z\sim 1.5$.
\label{madau.plot}}
\end{center}}

\subsection{Cosmic Star Formation History}

As pointed out by others (e.g., {Larson} 2005; {Hopkins} \& {Beacom} 2006; {Tumlinson} 2007; {Fardal} {et~al.} 2007),
an evolving IMF has direct implications for the derived cosmic star
formation history. Essentially
all indicators of the star formation rate (such
as ultraviolet luminosity,
H$\alpha$ line luminosity, and infrared luminosity) measure
the effects of luminous O and B stars, which have masses
$\gtrsim 10$\,\msun. Inferred total star formation rates
therefore rely on an extrapolation of the IMF from $\sim
10$\,\msun\ down to $\sim 0.1$\,\msun\
(see, e.g., {Lilly} {et~al.} 1996; {Madau} {et~al.} 1996; {Madau}, {Pozzetti}, \& {Dickinson} 1998). As most of the
total mass is in low mass stars, changes to
the IMF in the mass range 0.1--1\,\msun\
will have virtually no effect on the observed star formation indicators
but  potentially large effects on the actual total stellar
mass that is produced.

The effects of an evolving IMF of the form of Eq.\
\ref{toy.eq} are shown in Fig.\ \ref{madau.plot}(a). 
The broken line shows the integral of Eq.\ \ref{mod.eq}, i.e.,
the integrated stellar mass for an increasing
value of $m_c$ with redshift.
The solid line shows the integrated
mass relative to that in stars with masses $>10$\,\msun,
as the energy output of
stars in this mass range drives the star formation
measurements. The effects are substantial at high redshift,
and at $z\sim 6$
the star formation rate could be overestimated by a
factor of 3--4 relative to $z=0$.

In Fig.\ \ref{madau.plot}(b) the correction of panel (a) is
applied to measurements of the star formation rate at a range
of redshifts. The data are a combination of the extensive
compilation of {Hopkins} (2004) and recent measurements at
high redshift by {Bouwens} {et~al.} (2007). Extinction-corrected
values were used from both studies. The data as reported
in the literature are shown in grey, and the grey curve shows
a simple fit to the points. Black points are the same data,
corrected
to the modified {Chabrier} (2003) IMF proposed in this paper.
This correction is a combination of two effects:
a constant offset to account for the
difference between a {Salpeter} (1955) IMF (which is
assumed in the quoted literature) and a {Chabrier} (2003)
IMF, and the redshift-dependent effect shown in Fig.\ \ref{madau.plot}(a).
As is clear from the black line,
the cosmic star formation history has a fairly well-defined peak
at $z\sim 1.5$ if the IMF depends on redshift.

\subsection{$M/L$ Ratios}
\label{mass.sec}

The influence of an evolving IMF on the $M/L$ ratios of galaxies
is complex, as several competing effects play a role.
The number of low mass stars with respect to high
mass stars is reduced, which lowers
the $M/L$ ratio as these stars contribute little to the integrated light.
However, for $m_c\gtrsim 0.4$\,\msun\ the number of
turn-off stars is also reduced (see Fig.\ \ref{chabrier.plot}),
and these stars dominate the light at
rest-frame optical wavelengths. As a result, the net effect
of an increased $m_c$ on the $M/L$ ratio is generally smaller than
on the star formation rate. A further complication is that the
turn-off mass can be similar to the characteristic mass
(e.g.,
in elliptical galaxies at low redshift where both values are
$\sim 1$\,\msun). This means that the effect on the $M/L$ ratio
is not a constant, but depends on the age of the population.
A final complication is the mass in stellar remnants, which
is a larger fraction of the total stellar mass for more top-heavy
IMFs. As discussed by {Maraston} (2005) and in \S\,\ref{absml.sec}
IMFs that have a powerlaw
slope of zero from $0.1-100$\,\msun\ imply completely remnant-dominated
mass functions at old ages, with unrealistically high $M/L$ ratios.

A correct treatment of these issues requires full stellar population
synthesis modeling, which is beyond the scope of the present paper.
Instead, we used simple stellar evolutionary tracks 
to estimate what the net outcome is of the
various competing mechanisms. The
Yale-Yonsei isochrones were used ({Yi}, {Kim}, \& {Demarque} 2003;
{Demarque} {et~al.} 2004).
Monte-Carlo simulations of 100,000
stars were generated, with Solar metallicity, a range of ages, and
a {Salpeter} (1955) IMF. Next, the mass function was resampled
to match Eq.\ \ref{mod.eq} for a grid of values of $m_c$. The total
$V$ band luminosity was determined by linearly adding the light
of individual stars. The mass in living stars was determined by integrating
Eq.\ \ref{mod.eq} up to the turn-off mass. The mass in remnants
was calculated by integrating Eq.\ \ref{mod.eq} from the turn-off
mass to 100\,\msun\ and multiplying by the fraction
of the initial mass that is retained in the form of black holes, neutron
stars, and white dwarfs. This (age-dependent)
fraction was determined in the same
way as described in {Bruzual} \& {Charlot} (2003).

The results are shown in Fig.\ \ref{mlmc.plot}. At young ages
the light is dominated by relatively massive stars, and the behavior
is similar to that of the implied star formation rate (see
Fig.\ \ref{madau.plot}a). At intermediate ages and high values of
$m_c$ the turn-off mass is similar to the characteristic mass, and
the effect on the luminosity is similar to the effect on the
mass. The net effect is a roughly constant $M/L_V$ ratio as
a function of $m_c$. At the oldest ages the IMF becomes remnant-dominated,
with $\sim 80$\,\% of the mass in remnants for an age of 10\,Gyr and
$m_c = 1.5$\,\msun. Therefore, the $M/L_V$ ratios are actually
{\em higher} than for a standard {Chabrier} (2003)
IMF wih $m_c=0.08$\,\msun,
approaching or even exceeding those implied
by a {Salpeter} (1955) IMF.
Although these results are somewhat uncertain due to the crude
nature of our modeling, the qualitative conclusion is that the
$M/L_V$ ratios of galaxies are not necessarily strongly
reduced by an evolving IMF, in contrast to their star
formation rates.

\vbox{
\begin{center}
\leavevmode
\hbox{%
\epsfxsize=8.5cm
\epsffile{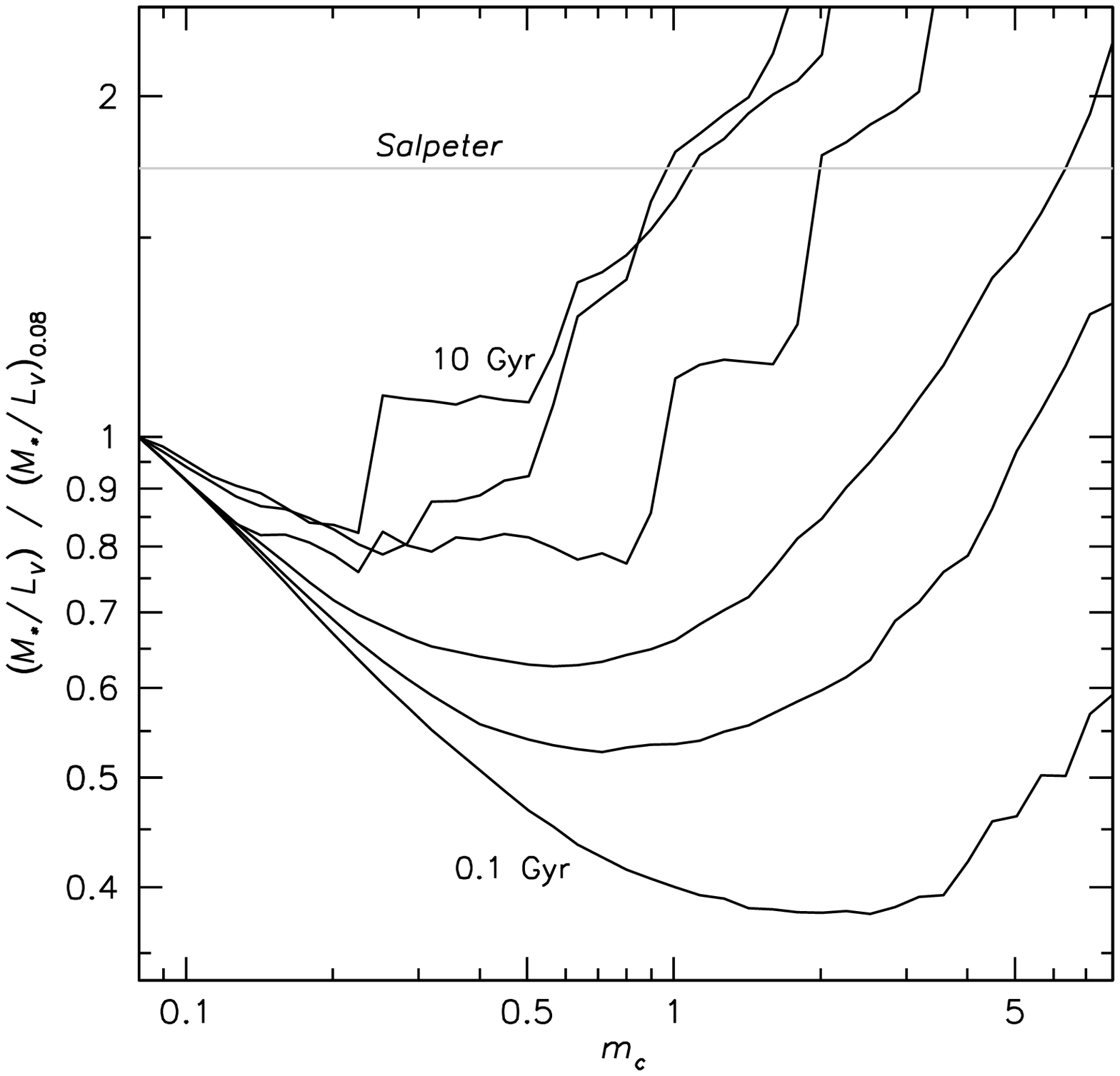}}
\figcaption{\small
Effects of a changing characteristic mass on the $M/L_V$ ratio
for stellar populations of ages 0.1, 0.3, 0.5, 3, 5, and 10 Gyr.
The results for higher ages are somewhat uncertain and noisy,
largely due to numerical effects. For young ages the $M/L$ ratio
steadily declines with increasing characteristic mass. The
behavior is more complex when
the characteristic mass is similar to the turn-off mass.
For $m_c \sim 1$\,\msun\ and old
ages the mass function becomes remnant-dominated,
and the $M/L$ ratios approach those implied by
a Salpeter (1955) IMF.
\label{mlmc.plot}}
\end{center}}

\subsection{Evolution of the Stellar Mass Density}

In Fig.\ \ref{mdens.plot} we show the effects of an evolving
IMF on the cosmic stellar mass density, in a similar way as
was done in Fig.\
\ref{madau.plot} for the cosmic star formation history.
Panel (a) shows the redshift dependence of the $M/L_V$ ratio,
with respect to the $M/L_V$ ratio implied by a standard
{Chabrier} (2003) IMF. To generate this relation it was assumed
that all galaxies start forming stars at $z=10$, and have a constant
star formation rate until the epoch of observation. The implied
mean age of galaxies at $z=6$ is $\sim 0.2$\,Gyr, and the mean
age of galaxies at $z=0$ is $\sim 6.5$\,Gyr. These values are
reasonable, and we note that
the results are not very sensitive to the details
of the star formation history of the galaxies.
This assumption
provides an estimate of the mean mass-weighted formation time of
the stars, and hence through Eq.\ \ref{toy.eq} the appropriate
value of $m_c$.
The relations between age, $m_c$, and $M/L_V$
ratio shown in Fig.\ \ref{mlmc.plot} are then used to estimate
the change in $M/L_V$ as a function of redshift.

\vbox{
\begin{center}
\leavevmode
\hbox{%
\epsfxsize=8.5cm
\epsffile{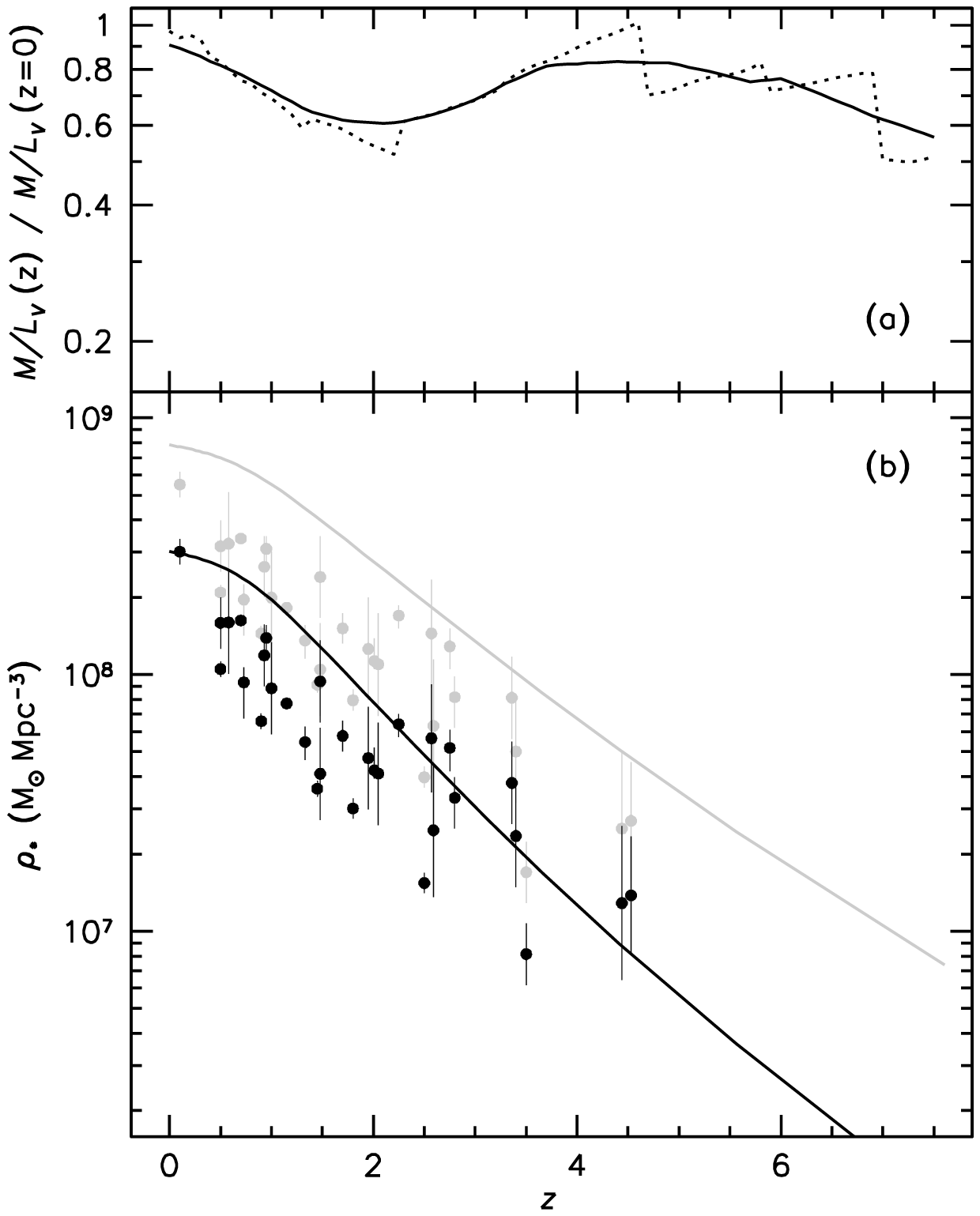}}
\figcaption{\small
Effects of an evolving IMF on the cosmic stellar
mass density. The broken line in
(a) shows the effect on the $M/L_V$
ratio with respect to a standard Chabrier (2003) IMF. 
Discontinuities are due to the discrete grid of ages
used in the computation; the solid line is a smoothed version
of the dotted line. (b) Measurements of the mass
density from Cole et al.\ (2001),
Dickinson et al.\ (2003), Drory et al.\ (2005), Rudnick et al.\
(2006), and Fontana et al.\ (2006). Grey symbols are
for a Salpeter (1955) IMF, and black symbols are
for the evolving IMF proposed in this paper.
The grey and black lines show the evolution implied by the
evolution of the star formation rate shown in Fig.\
\ref{madau.plot}. The black line
is a somewhat better fit to the black points than the
grey line is to the grey points.
\label{mdens.plot}}
\end{center}}

Going up in redshift, the $M/L_V$ ratio first declines due to
the increase in $m_c$ from $\sim 0.1$\,\msun\ to $\sim 0.4$\,\msun.
The $M/L_V$ ratio reaches a minimum at $z\sim 2$, and then
increases  as the characteristic mass becomes similar
to the turn-off mass. Beyond $z\sim 4$ the $M/L$ ratio
once again decreases, as the galaxies become younger and
the turn-off mass increases rapidly.
Measurements of the evolution of the mass density
from {Cole} {et~al.} (2001), {Dickinson} {et~al.} (2003), {Drory} {et~al.} (2005),
{Rudnick} {et~al.} (2006), and
{Fontana} {et~al.} (2006) are shown in panel (b).
Errorbars are taken from the literature sources, and
typically do not include systematic uncertainties.
Grey points show the literature values, determined assuming
a {Salpeter} (1955) IMF, and black points show the corrected
values.  The correction is
a constant offset (to account for the difference between
a {Salpeter} (1955) IMF and a standard
{Chabrier} (2003) IMF)
in addition to the relation shown in panel (a).

The grey and black lines show the evolution of the stellar
mass density as implied by the observed evolution of the
star formation rate shown in Fig.\ \ref{madau.plot}.
Mass loss was taken into account using the same scheme as
employed by {Bruzual} \& {Charlot} (2003). Note that
mass loss is a larger effect for Chabrier-like IMFs than
for a {Salpeter} (1955) IMF, as a larger fraction of the
total mass is in high mass stars.

The median difference between the data points and the curve
is a factor of 2.3 for a non-evolving IMF, and a factor
of 1.7 for the evolving Chabrier-like IMF.
Considering that this type of comparison has many systematic
uncertainties quite independent of the IMF ({Hopkins} \& {Beacom} 2006; {Fardal} {et~al.} 2007), the fact that the two
independent measures of the build-up of stellar mass in the
Universe agree to within a factor of $\sim 2$ can be considered
a success. In any case, the discrepancy is smaller for the
evolving IMF than for a non-evolving IMF, because
the star formation rate is reduced by a larger fraction
than the mass density. Our evolving IMF has a qualitatively
similar effect
as the ``paunchy'' IMF with an increased contribution from
stars around 1.5--4\,\msun\ proposed by {Fardal} {et~al.} (2007),
which was specifically designed to give better agreement
between the star formation
history of the Universe (as implied by the extragalactic background
radiation) and the observed evolution of the mass density.

\section{Summary and Conclusions}
\label{conclusion.sec}

This paper compares the color evolution of massive cluster galaxies
to their luminosity evolution, with the aim of constraining the
form of the IMF at the time when the stars in these galaxies
were formed. It is found that the evolution
of the rest-frame $U-V$ color
is not consistent with the previously determined
evolution of the rest-frame $M/L_B$ ratio, unless the IMF slope
is significantly flatter than the Salpeter value around 1\,\msun.
For standard IMFs with a slope of 1.3 at $m\geq 1$\,\msun\
the luminosity evolution is too fast for the measured color
evolution, and the implied stellar ages derived from $M/L$
evolution and color evolution are not consistent with each
other. The only models that are able to fit the color evolution
and the luminosity evolution simultaneously have IMF slopes of
$\sim 0$ around 1\,\msun\ and mean luminosity-weighted
stellar formation redshifts of $\sim 4$ (for Solar metallicity).

This result is somewhat uncertain, as the currently available
sample of cluster galaxies with accurate rest-frame $U-V$ colors
and dynamical masses is somewhat limited and there are many
systematic effects which may play a role. In particular,
it is an open question
whether stellar population synthesis models are able to predict
color evolution with the required accuracy. The commonly used
{Bruzual} \& {Charlot} (2003) and {Maraston} (2005) models give broadly similar
answers, but that may be because they share many of the same
assumptions. 

As discussed in \S\,\ref{comp.sec} the higher stellar ages implied
by a flat IMF are consistent with many other studies, which lends
some credibility to the results presented here.
Of particular importance is the agreement with the data
on Balmer line strengths of {Kelson} {et~al.} (2001), as they do not
suffer from the same
systematic uncertainties as the color data.
Formation redshifts substantially larger than two also fit
more comfortably with
the direct detection of old galaxies at
high redshifts. A firm independent measurement of the
star formation epoch of massive cluster galaxies, combined
with a better understanding of selection effects at high
redshift, would
leave the IMF as the only free parameter and greatly simplify
the problem.

The implications discussed in \S\,\ref{imply.sec} are obviously
somewhat speculative. Although the interpretation in terms of an
evolving characteristic mass is physically plausible according
to some models (e.g., {Larson} 2005), many other forms of the
IMF are consistent with the data. The observations described in this
paper are only sensitive to a narrow mass range near 1\,\msun, and
the IMF proposed in Eq.\ \ref{mod.eq} represents a very substantial
extrapolation. This is illustrated in Fig.\ \ref{summary.plot}:
both the top-heavy IMF (red line) and the ``bottom-light'' IMF
(green line) are consistent with the data presented in this paper.
The main reason for preferring the bottom-light IMF over a
top-heavy form is that the absolute
$M/L$ ratios of galaxies are within a reasonable range. As shown in
Fig.\ \ref{mlmc.plot} the $M/L_V$ ratios are similar to those implied
by a standard {Chabrier} (2003) IMF, which means that they are
consistent with dynamical measurements at $z=0$ ({Cappellari} {et~al.} 2006).

\vbox{
\begin{center}
\leavevmode
\hbox{%
\epsfxsize=8.5cm
\epsffile{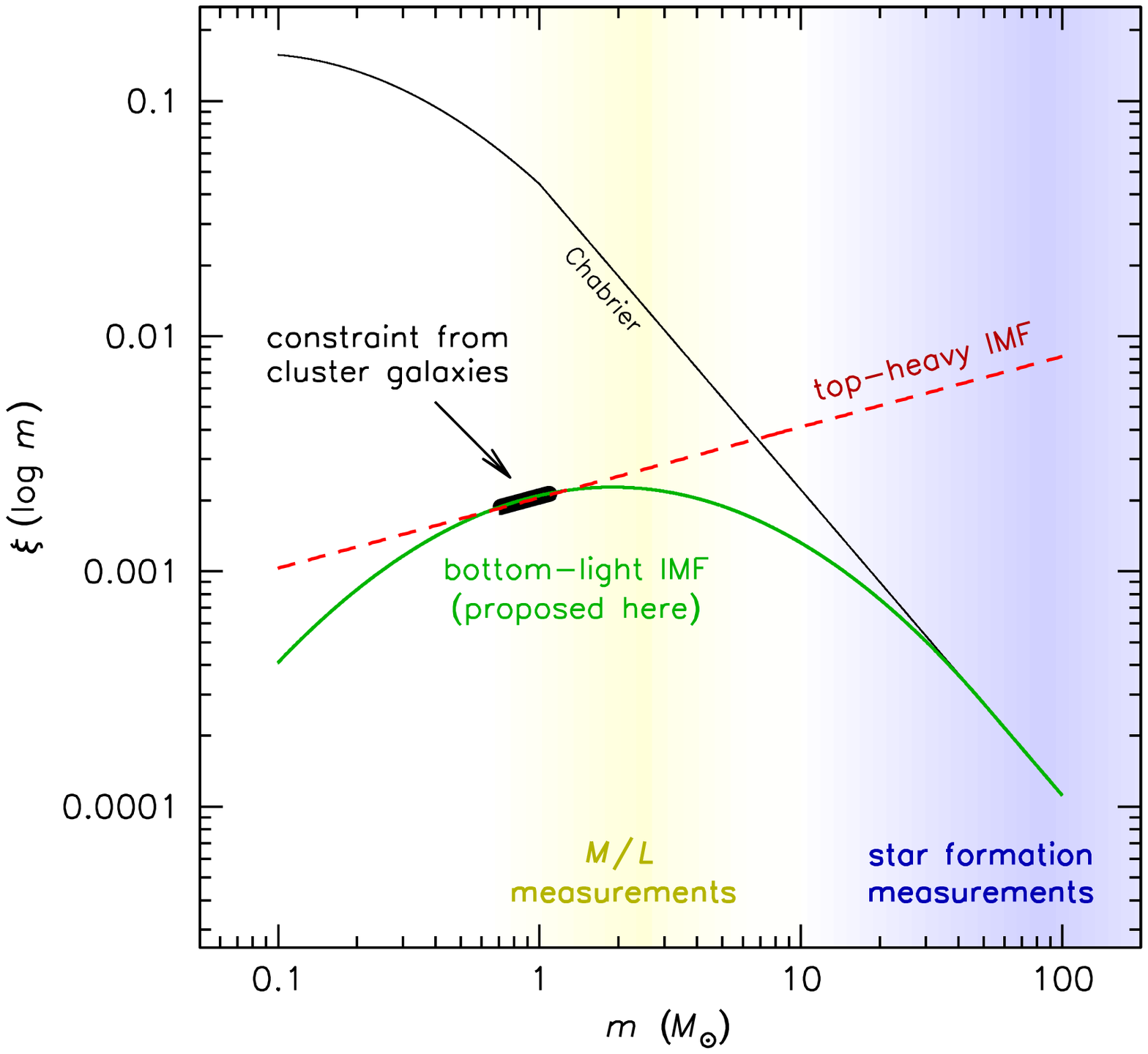}}
\figcaption{\small
Illustration of the key results of this paper. In
\S\,4 the slope of the IMF of massive cluster galaxies
$x$ is estimated to be approximately $-0.3$
in a narrow region around 1\,\msun\ (thick black line).
The red dashed line and the green solid line show two
possible interpretations: a global change of the slope
of the IMF (with respect
to a standard Chabrier or Salpeter IMF)
at all masses (red dashed line), or a change
in the characteristic mass (solid green line). The ``top-heavy''
interpretation is inconsistent with the dynamical
$M/L$ ratios of nearby elliptical galaxies (see \S\,\ref{absml.sec}),
whereas the ``bottom-light'' interpretation is consistent with
all data that we are aware of. The blue and yellow areas illustrate
that stars with masses $\gtrsim 10$\,\msun\ drive star formation
measurements, whereas stars with masses $1-5$\,\msun\ drive $M/L$
measurements (see \S\,7.3-7.5).
\label{summary.plot}}
\end{center}}

An ``unintended'' effect of
an evolving IMF of the form proposed here is that it
reduces the discrepancy between the observed
stellar mass density and the density
implied by the cosmic star formation history.
This result is in excellent (albeit qualitative)
agreement with several other
recent studies (e.g., Hopkins \& Beacom 2006, Fardal et al.\ 2007,
P\'erez-Gonz\'alez et al.\ 2007; Wilkins, Trentham, \&
Hopkins 2007; Dav\'e 2007; see also, e.g., Fields 1999).
The differences
between a non-evolving IMF and an evolving IMF are fairly
large at $z\sim 4$ (as the effect on the star formation rate is
strong and the effect on $M/L$ ratios is weak at that redshift),
and it will be interesting to see where future
measurements of
the mass density at high redshift will fall in Fig.\ \ref{mdens.plot}.
We note that
the effects on the $M/L_V$ ratios are somewhat uncertain, as they
rely on rather rudimentary stellar population synthesis modeling.
The effects on star formation rates are more robust, and
suggest that the cosmic star formation rate peaked at $z\sim 1.5$.

This paper adds to previous theoretically and observationally
motivated suggestions that the IMF may evolve with redshift
(e.g., Worthey et al.\ 1992; Larson 1998, 2005;
Fields 1999; Blain et al.\ 1999a; Baugh et al.\ 2005;
Stanway et al.\ 2005; Hopkins \& Beacom 2006; Tumlinson
2007; Lacey et al.\ 2007;
Fardal et al.\ 2007; P\'erez-Gonz\'alez et al.\ 2007). Although
these studies vary greatly in their parameterization of IMF
evolution and the range of stellar masses that are considered,
they all suggest that the ratio of high-mass stars to low-mass
stars was higher in the past.
It should be pointed out that most of these papers
invoke a change in the IMF as a ``last resort'' possibility, to
explain data that are otherwise difficult to interpret. In the
present study a different approach was followed, in that we
set out with the specific purpose of constraining the slope of
the IMF. An advantage of the applied method is that it is fairly
direct, as the rate of luminosity evolution is
determined by the number of stars as a function of mass.
Disadvantages are that it is only sensitive to a very limited
mass range (see Fig.\ \ref{summary.plot}); that it relies on
stellar population synthesis models, which are not well calibrated
in the relevant parameter range; that the progenitors of
early-type galaxies may not be representative for the general
population of high redshift galaxies; and that the currently
available data are somewhat limited.

Accepting the possibility of an evolving characteristic mass, it
is interesting to speculate what could be the cause, or causes.
The proximate cause may well be a higher temperature in molecular
clouds at high redshift, which would raise the Jeans mass and
could inhibit the formation of low mass stars ({Larson} 1998, 2005).
The ultimate cause could be the higher temperature of the cosmic
microwave background, the fact that star formation tends to
proceed in more extreme environments at higher redshift, or
a combination. Available information on IR-bright galaxies
suggests that dust temperatures in star burst galaxies
are of order $30-40$\,K ({Dunne} {et~al.} 2000; {Chapman} {et~al.} 2005),
and hence exceed the CMB temperature for all relevant redshifts.
However, it is as yet unclear what fraction of the total star
formation has taken place in these extreme environments
(see, e.g., {Reddy} {et~al.} 2007).

The analysis presented here can be improved in various ways.
The number of clusters with accurate rest-frame $U-V$ is currently
smaller than the number of clusters with accurate $M/L_B$
measurements, and this can be remedied by obtaining accurate
(space-based) photometry in well-chosen filters
of the remaining clusters in the vv07 sample.
It is also important to measure
the evolution in a redder rest-frame color, such as
$V-I$. Redder color suffer less from possible contributions
of hot stars, and their evolution is
probably somewhat better calibrated in stellar population synthesis
models. This requires very accurate photometry in the
near-infrared, which should be possible with WFC3 on HST.
On the modeling side, it would be helpful to implement more
variations of the IMF in stellar population synthesis codes than the
standard Salpeter, Kroupa, and Chabrier forms. Ultimately it
may be fruitful (or prove necessary) to have the characteristic mass, or
some other parameter describing the form of the IMF, as one of
the ``standard'' parameters in these models, on a par with
the age and metallicity.

\begin{acknowledgements}
The author acknowledges very helpful discussions with 
Richard Larson, Marijn Franx, Charles Bailyn,
Arjen van der Wel, and Bob Zinn. Claudia Maraston kindly provided
model predictions for different slopes of the IMF. Daniel
McIntosh shared unpublished data on low redshift clusters.
Comments of the anonymous referee
significantly improved the presentation.
Support from NSF CAREER grant AST 04-49678 and NASA LTSA grant
NNG04GE12G is gratefully acknowledged.
\end{acknowledgements}


\begin{appendix}

\section{Aperture Corrections}

For five of the seven distant clusters discussed in this Paper
colors were measured directly from HST images
in apertures of fixed physical size. For
each cluster this size was chosen to match an $11\arcsec$
diameter aperture at the distance of the Coma cluster, allowing
a direct comparison to data from BLE92 and to {McIntosh} {et~al.} (2005).
For the clusters MS\,1054--03 ($z=0.831$) and RX\,J0152--13
($z=0.837$) color measurements were taken from {Blakeslee} {et~al.} (2006),
who used apertures scaled to contain 50\,\% of each galaxy's light.
In this Section we give the corrections that were applied
to bring the {Blakeslee} {et~al.} (2006) data onto our system, and test
these corrections using data from {McIntosh} {et~al.} (2005).

The reason why aperture corrections need to be applied even when
data are corrected for seeing effects is that early-type galaxies
have color gradients. These gradients do not vary much from
galaxy to galaxy, when expressed as
$\Delta (F_1-F_2)/\Delta(\log r)$, with $F_1$ and $F_2$ the relevant
filters and $\log r$ the logarithm of the circularized radial
distance to the center of the galaxy (see, e.g., {Franx}, {Illingworth}, \&  {Heckman} 1989; {Franx} \& {Illingworth} 1990; {Peletier} {et~al.} 1990). The measured color within an aperture
$r_{\rm ap}$ is not simply the integral of the color gradient,
as the radial dependence of the luminosity has to be taken into
account as well.

We determine the expected dependence of color
on aperture size numerically, assuming an $r^{1/4}$ law for the
radial luminosity dependence. The solid line in
Fig.\ \ref{colgrad.plot} shows
the expected relation for a color gradient of $-0.15$,
appropriate for $U-V$ (e.g., {Peletier} {et~al.} 1990).
The long-dashed line is a polynomial fit to the numerically-derived
relation, of the form
\begin{equation}
\label{colgrad.eq}
(F_1-F_2)_{\rm ap} = (F_1-F_2)_{\rm eff} + 6.67 \frac{\Delta (F_1-F_2)}
{\Delta \log r}
\left( - 0.118 \log \left( 2 r_e / D_{\rm ap} \right) -
0.042 \left( \log \left( 2 r_e/D_{\rm ap} \right) \right)^2 \right),
\end{equation}
with $(F_1-F_2)_{\rm ap}$ the color measured through an aperture of
diameter $D_{\rm ap}$, $(F_1-F_2)_{\rm eff}$ the color measured through
an aperture containing 50\,\% of the galaxy's luminosity, $r_e$
the half-light radius, and $\Delta (F_1-F_2) / \Delta \log r = -0.15$ the
color gradient in mag per dex. The long-dashed line is a very good
match to the (exact) solid line. For reference, short-dashed lines
show the behavior of Eq.\ \ref{colgrad.eq} for color gradients of
$-0.10$ and $-0.20$. The data points show binned data from
{McIntosh} {et~al.} (2005), who measured $U-V$ colors of three nearby
clusters in apertures of fixed size and in apertures of
diameter $2r_e$. The data follow the predicted relation very closely,
with an rms of $\sim 0.01$ and no apparent systematic deviations.

\vbox{
\begin{center}
\leavevmode
\hbox{%
\epsfxsize=8.5cm
\epsffile{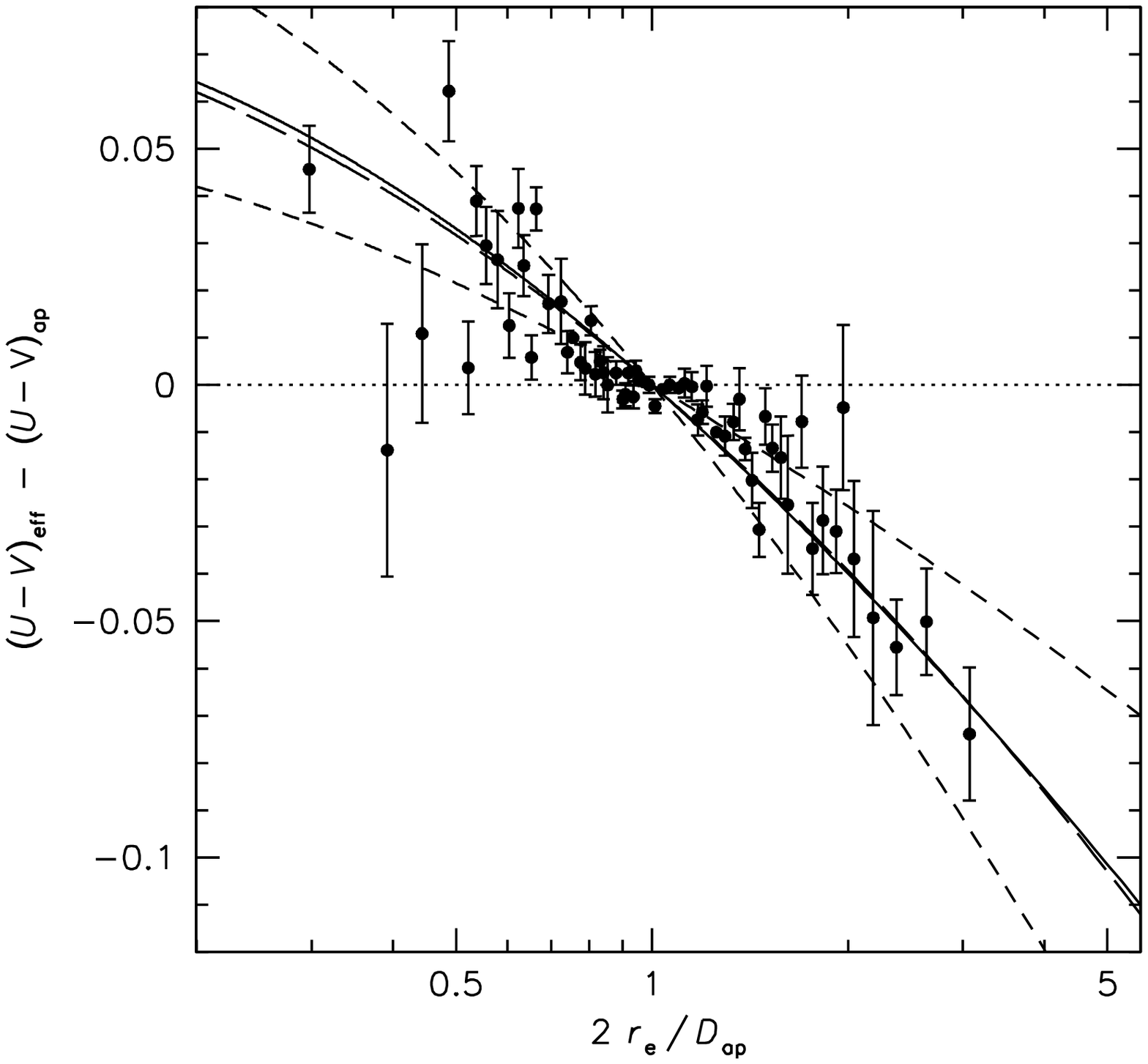}}
\figcaption{\small
Effect of color gradients on the measured color within an
aperture. The solid line shows the expected difference between the
color measured within aperture $D_{\rm ap}$ and the color measured
within an aperture containing 50\,\% of the galaxy's light, assuming
an $r^{1/4}$ law and a color gradient of $-0.15$ mag per dex. Datapoints
are from McIntosh et al.\ (2005); they follow the predicted curve
very well. Broken lines are simple polynomial fits to the predicted
evolution for color gradients of $-0.10$, $-0.15$, and $-0.20$.
\label{colgrad.plot}}
\end{center}}

The datapoints from {Blakeslee} {et~al.} (2006) were corrected to our fiducial
aperture size by calculating $r_e/D_{\rm ap}$ for each galaxy (using
the information supplied in Blakeslee et al.) and applying Eq.\
\ref{colgrad.eq}. As on average $D_{\rm ap} \sim 2r_e$ for our
aperture size the correction is only $\sim 0.015$ in the mean
for both clusters, with an
object-to-object scatter of $\sim 0.03$. Galaxies requiring a correction
$>0.05$ or $<-0.05$ were removed from the sample; this step
slightly reduces the scatter
in the color-mass relations of the clusters but otherwise
has a negligible effect on our results.

\end{appendix}


\begin{references}

\reference{}{Abel}, T., {Bryan}, G.~L., \& {Norman}, M.~L. 2002, Science, 295, 93

\reference{}{Almeida}, C., {Baugh}, C.~M., \& {Lacey}, C.~G. 2007, \mnras, 376, 1711

\reference{}{Andreon}, S. 2003, \aap, 409, 37

\reference{}{Bailyn}, C.~D. 1995, \araa, 33, 133

\reference{}{Baugh}, C.~M., {Lacey}, C.~G., {Frenk}, C.~S., {Granato}, G.~L., {Silva}, L.,  {Bressan}, A., {Benson}, A.~J., \& {Cole}, S. 2005, \mnras, 356, 1191

\reference{}{Beers}, T.~C., {Flynn}, K., \& {Gebhardt}, K. 1990, \aj, 100, 32

\reference{}{Bell}, E.~F., {Naab}, T., {McIntosh}, D.~H., {Somerville}, R.~S., {Caldwell},  J.~A.~R., {Barden}, M., {Wolf}, C., {Rix}, H.-W., {et al.} 2006, \apj, 640, 241

\reference{}{Bessell}, M.~S. 1990, \pasp, 102, 1181

\reference{}{Blain}, A.~W., {Jameson}, A., {Smail}, I., {Longair}, M.~S., {Kneib}, J.-P.,  \& {Ivison}, R.~J. 1999a, \mnras, 309, 715

\reference{}{Blain}, A.~W., {Smail}, I., {Ivison}, R.~J., \& {Kneib}, J.-P.  1999b, \mnras, 302, 632

\reference{}{Blakeslee}, J.~P., {Holden}, B.~P., {Franx}, M., {Rosati}, P., {Bouwens},  R.~J., {Demarco}, R., {Ford}, H.~C., {Homeier}, N.~L., {et al.} 2006, \apj, 644, 30

\reference{}{Bouwens}, R.~J., {Illingworth}, G.~D., {Franx}, M., \& {Ford}, H. 2007, \apj, in press (arXiv:0707.2080)

\reference{}{Bower}, R.~G., {Lucey}, J.~R., \& {Ellis}, R.~S. 1992a, \mnras,  254, 601

\reference{}---. 1992b, \mnras, 254, 589

\reference{}{Bromm}, V., {Coppi}, P.~S., \& {Larson}, R.~B. 2002, \apj, 564, 23

\reference{}{Bruzual}, G. 2007, in IAU Symposium No.\ 241, Stellar
populations as building blocks of galaxies, ed.\ A.\ Vazdekis, \&
R.\ Peletier, in press (astro-ph/0703052)

\reference{}{Bruzual}, G. \& {Charlot}, S. 2003, \mnras, 344, 1000

\reference{}{Burstein}, D., {Bertola}, F., {Buson}, L.~M., {Faber}, S.~M., \& {Lauer},  T.~R. 1988, \apj, 328, 440

\reference{}{Calzetti}, D. 1997, \aj, 113, 162

\reference{}{Cappellari}, M., {Bacon}, R., {Bureau}, M., {Damen}, M.~C., {Davies}, R.~L.,  {de Zeeuw}, P.~T., {Emsellem}, E., {Falc{\'o}n-Barroso}, J., {et al.} 2006, \mnras, 366, 1126

\reference{}{Chabrier}, G. 2003, \pasp, 115, 763

\reference{}{Chapman}, S.~C., {Blain}, A.~W., {Ivison}, R.~J., \& {Smail}, I.~R. 2003,  \nat, 422, 695

\reference{}{Chapman}, S.~C., {Blain}, A.~W., {Smail}, I., \& {Ivison}, R.~J. 2005, \apj,  622, 772

\reference{}{Cole}, S., {Norberg}, P., {Baugh}, C.~M., {Frenk}, C.~S., {Bland-Hawthorn},  J., {Bridges}, T., {Cannon}, R., {Colless}, M., {et al.} 2001, \mnras, 326, 255

\reference{}{Coleman}, G.~D., {Wu}, C.-C., \& {Weedman}, D.~W. 1980, \apjs, 43, 393

\reference{}{Daddi}, E., {Renzini}, A., {Pirzkal}, N., {Cimatti}, A., {Malhotra}, S.,  {Stiavelli}, M., {Xu}, C., {Pasquali}, A., {et al.} 2005, \apj, 626, 680

\reference{}Dav\'e, R. 2007, \mnras, submitted (astro-ph/0710.0381)

\reference{}{de Marchi}, G., {Paresce}, F., \& {Portegies Zwart}, S. 2005, in Astrophysics  and Space Science Library, Vol. 327, The Initial Mass Function 50 Years  Later, ed. E.~{Corbelli}, F.~{Palla}, \& H.~{Zinnecker}, 77

\reference{}{Demarque}, P., {Woo}, J.-H., {Kim}, Y.-C., \& {Yi}, S.~K. 2004, \apjs, 155,  667

\reference{}{Dickinson}, M., {Papovich}, C., {Ferguson}, H.~C., \& {Budav{\'a}ri}, T. 2003,  \apj, 587, 25

\reference{}{Djorgovski}, S. \& {Davis}, M. 1987, \apj, 313, 59

\reference{}{Dorman}, B., {Rood}, R.~T., \& {O'Connell}, R.~W. 1993, \apj, 419, 596

\reference{}{Dressler}, A. \& {Gunn}, J.~E. 1983, \apj, 270, 7

\reference{}{Drory}, N., {Salvato}, M., {Gabasch}, A., {Bender}, R., {Hopp}, U., {Feulner},  G., \& {Pannella}, M. 2005, \apjl, 619, L131

\reference{}{Dunne}, L., {Eales}, S., {Edmunds}, M., {Ivison}, R., {Alexander}, P., \&  {Clements}, D.~L. 2000, \mnras, 315, 115

\reference{}{Dwek}, E., {Arendt}, R.~G., {Hauser}, M.~G., {Fixsen}, D., {Kelsall}, T.,  {Leisawitz}, D., {Pei}, Y.~C., {Wright}, E.~L., {et al.} 1998, \apj, 508, 106

\reference{}{Eisenhardt}, P.~R., {De Propris}, R., {Gonzalez}, A.~H., {Stanford}, S.~A.,  {Wang}, M., \& {Dickinson}, M. 2007, \apjs, 169, 225

\reference{}{Ellis}, R.~S., {Smail}, I., {Dressler}, A., {Couch}, W.~J., {Oemler}, A.~J.,  {Butcher}, H., \& {Sharples}, R.~M. 1997, \apj, 483, 582

\reference{}{Evans}, II, N.~J., {Rawlings}, J.~M.~C., {Shirley}, Y.~L., \& {Mundy}, L.~G.  2001, \apj, 557, 193

\reference{}{Fardal}, M.~A., {Katz}, N., {Weinberg}, D.~H., \& {Dav{\'e}}, R. 2007, \mnras,  379, 985

\reference{}Fields, B.~D.\ 1999, \apj, 515, 603

\reference{}{Figer}, D.~F., {Kim}, S.~S., {Morris}, M., {Serabyn}, E., {Rich}, R.~M., \&  {McLean}, I.~S. 1999, \apj, 525, 750

\reference{}{Fontana}, A., {Salimbeni}, S., {Grazian}, A., {Giallongo}, E., {Pentericci},  L., {Nonino}, M., {Fontanot}, F., {Menci}, N., {et al.} 2006, \aap, 459, 745

\reference{}{Franx}, M. 1993, \pasp, 105, 1058

\reference{}{Franx}, M. \& {Illingworth}, G. 1990, \apjl, 359, L41

\reference{}{Franx}, M., {Illingworth}, G., \& {Heckman}, T. 1989, \aj, 98, 538

\reference{}{Franx}, M., {Labb{\' e}}, I., {Rudnick}, G., {van Dokkum}, P.~G., {Daddi}, E.,  {F{\" o}rster Schreiber}, N.~M., {Moorwood}, A., {Rix}, H., {et al.} 2003, \apjl, 587, L79

\reference{}{Gavazzi}, R., {Treu}, T., {Rhodes}, J.~D., {Koopmans}, L.~V.~E., {Bolton},  A.~S., {Burles}, S., {Massey}, R., \& {Moustakas}, L.~A. 2007,
\apj, 667, 176

\reference{}{Goudfrooij}, P., {de Jong}, T., {Hansen}, L., \& {Norgaard-Nielsen}, H.~U.  1994, \mnras, 271, 833

\reference{}{Gratton}, R.~G., {Bragaglia}, A., {Carretta}, E., {Clementini}, G.,  {Desidera}, S., {Grundahl}, F., \& {Lucatello}, S. 2003, \aap, 408, 529

\reference{} Harayama, Y., Eisenhauer, F., \& Martins, F. 2007, \apj,
 in press (astro-ph/0710.2882)

\reference{}{Holden}, B.~P., {Franx}, M., {Illingworth}, G.~D., {Postman}, M., {Blakeslee},  J.~P., {Homeier}, N., {Demarco}, R., {Ford}, H.~C., {et al.} 2006, \apjl, 642, L123

\reference{}{Holden}, B.~P., {Illingworth}, G.~D., {Franx}, M., {Blakeslee}, J.~P.,  {Postman}, M., {Kelson}, D.~D., {van der Wel}, A., {Demarco}, R., {et al.} 2007, \apj, in press (arXiv:0707.2782)

\reference{}{Holden}, B.~P., {Stanford}, S.~A., {Eisenhardt}, P., \& {Dickinson}, M. 2004,  \aj, 127, 2484

\reference{}{Hopkins}, A.~M. 2004, \apj, 615, 209

\reference{}{Hopkins}, A.~M. \& {Beacom}, J.~F. 2006, \apj, 651, 142

\reference{}{Jappsen}, A.-K., {Klessen}, R.~S., {Larson}, R.~B., {Li}, Y., \& {Mac Low},  M.-M. 2005, \aap, 435, 611

\reference{}{J\o{}rgensen}, I., {Chiboucas}, K., {Flint}, K., {Bergmann}, M., {Barr}, J.,  \& {Davies}, R. 2006, \apjl, 639, L9

\reference{}J\o{}rgensen, I., {Franx}, M., \& {Kj\ae{}rgaard}, P. 1995a, \mnras, 273, 1097

\reference{}---. 1995b, \mnras, 276, 1341

\reference{}---. 1996, \mnras, 280, 167

\reference{}{Joshi}, K.~J., {Nave}, C.~P., \& {Rasio}, F.~A. 2001, \apj, 550, 691

\reference{}{Kelson}, D.~D., {Illingworth}, G.~D., {Franx}, M., \& {van Dokkum}, P.~G.  2001, \apjl, 552, L17

\reference{}---. 2006, \apj, 653, 159

\reference{}{Kim}, S.~S., {Figer}, D.~F., {Kudritzki}, R.~P., \& {Najarro}, F. 2006, \apjl,  653, L113

\reference{}{Kodama}, T., {Arimoto}, N., {Barger}, A.~J., \& {Arag'on-Salamanca}, A. 1998,  \aap, 334, 99

\reference{}{Kodama}, T., {Tanaka}, I., {Kajisawa}, M., {Kurk}, J., {Venemans}, B., {De  Breuck}, C., {Vernet}, J., \& {Lidman}, C. 2007, \mnras, 377, 1717

\reference{}{Koo}, D.~C., {Datta}, S., {Willmer}, C.~N.~A., {Simard}, L., {Tran}, K.-V., \&  {Im}, M. 2005, \apjl, 634, L5

\reference{}{Koopmans}, L.~V.~E., {Treu}, T., {Bolton}, A.~S., {Burles}, S., \&  {Moustakas}, L.~A. 2006, \apj, 649, 599

\reference{}{Kriek}, M., {van Dokkum}, P., {Franx}, M., Quadri, R.,
Gawiser, E., Herrera, D., Illingworth, G.~D., Labb\'e, I., et al.\ 2006,
\apj, 649, L71

\reference{}{Krist}, J. 1995, in ASP Conf. Ser. 77: Astronomical Data Analysis Software and  Systems IV, Vol.~4, 349

\reference{}{Kroupa}, P. 2001, \mnras, 322, 231

\reference{}---. 2002, Science, 295, 82

\reference{}{Labb{\'e}}, I., {Huang}, J., {Franx}, M., {Rudnick}, G., {Barmby}, P.,  {Daddi}, E., {van Dokkum}, P.~G., {Fazio}, G.~G., {et al.} 2005, \apjl, 624, L81

\reference{}{Lacey}, C.~G., {Baugh}, C.~M., {Frenk}, C.~S., {Silva}, L., {Granato}, G.~L.,  \& {Bressan}, A. 2007, \mnras, submitted (arXiv:0704.1562)

\reference{}{Landolt}, A.~U. 1992, \aj, 104, 340

\reference{}{Larson}, R.~B. 1985, \mnras, 214, 379

\reference{}---. 1998, \mnras, 301, 569

\reference{}---. 2003, Reports of Progress in Physics, 66, 1651

\reference{}---. 2005, \mnras, 359, 211

\reference{}{Lilly}, S.~J., {Le Fevre}, O., {Hammer}, F., \& {Crampton}, D. 1996, \apjl,  460, L1+

\reference{}{Madau}, P., {Ferguson}, H.~C., {Dickinson}, M.~E., {Giavalisco}, M.,  {Steidel}, C.~C., \& {Fruchter}, A. 1996, \mnras, 283, 1388

\reference{}{Madau}, P., {Pozzetti}, L., \& {Dickinson}, M. 1998, \apj, 498, 106

\reference{}{Makino}, J. \& {Hut}, P. 1997, \apj, 481, 83

\reference{}{Mamon}, G.~A. \& {{\L}okas}, E.~L. 2005, \mnras, 363, 705

\reference{}{Maness}, H., {Martins}, F., {Trippe}, S., {Genzel}, R., {Graham}, J.~R.,  {Sheehy}, C., {Salaris}, M., {Gillessen}, S., {et al.} 2007,
\apj, in press (arXiv:0707.2382)

\reference{}{Maraston}, C. 1998, \mnras, 300, 872

\reference{}---. 2005, \mnras, 362, 799

\reference{}{Maraston}, C., {Daddi}, E., {Renzini}, A., {Cimatti}, A., {Dickinson}, M.,  {Papovich}, C., {Pasquali}, A., \& {Pirzkal}, N. 2006, \apj, 652, 85

\reference{}{Mateo}, M., {Harris}, H.~C., {Nemec}, J., \& {Olszewski}, E.~W. 1990, \aj,  100, 469

\reference{}{McCrady}, N., {Gilbert}, A.~M., \& {Graham}, J.~R. 2003, \apj, 596, 240

\reference{}{McCrady}, N., {Graham}, J.~R., \& {Vacca}, W.~D. 2005, \apj, 621, 278

\reference{}{McIntosh}, D.~H., {Zabludoff}, A.~I., {Rix}, H.-W., \& {Caldwell}, N. 2005,  \apj, 619, 193

\reference{}{McKee}, C.~F. \& {Ostriker}, E.~C. 2007, ARA\&{}A, 45, 565

\reference{}{Miller}, G.~E. \& {Scalo}, J.~M. 1979, \apjs, 41, 513

\reference{}{Nagamine}, K., {Cen}, R., {Hernquist}, L., {Ostriker}, J.~P., \& {Springel},  V. 2005, \apj, 627, 608

\reference{}{Padoan}, P. \& {Nordlund}, {\AA}. 2002, \apj, 576, 870

\reference{}{Paresce}, F. \& {De Marchi}, G. 2000, \apj, 534, 870

\reference{}{Peletier}, R.~F. 1989, PhD thesis,
University of Groningen, The Netherlands

\reference{}{Peletier}, R.~F., {Davies}, R.~L., {Illingworth}, G.~D., {Davis}, L.~E., \&  {Cawson}, M. 1990, \aj, 100, 1091

\reference{} P\'erez-Gonz\'alez, P.~G., Rieke, G.~H., Villar, V., Barro, G.,
Blaylock, M., Egami, E., Gallego, J., Gil de Paz, A., et al.\ 2007,
\apj, in press (astro-ph/0709.1354)

\reference{}{Pettini}, M., {Steidel}, C.~C., {Adelberger}, K.~L., {Dickinson}, M., \&  {Giavalisco}, M. 2000, \apj, 528, 96

\reference{}{Quadri}, R., {van Dokkum}, P., {Gawiser}, E., {Franx}, M., {Marchesini}, D.,  {Lira}, P., {Rudnick}, G., {Herrera}, D., {et al.} 2007, \apj, 654, 138

\reference{}{Reddy}, N.~A., {Steidel}, C.~C., {Pettini}, M., {Adelberger}, K.~L.,  {Shapley}, A.~E., {Erb}, D.~K., \& {Dickinson}, M. 2007,
\apj, in press (arXiv:0706.4091)

\reference{}{Rich}, R.~M., {Sosin}, C., {Djorgovski}, S.~G., {Piotto}, G., {King}, I.~R.,  {Renzini}, A., {Phinney}, E.~S., {Dorman}, B., {et al.} 1997, \apjl, 484, L25+

\reference{}{Rieke}, G.~H., {Loken}, K., {Rieke}, M.~J., \& {Tamblyn}, P. 1993, \apj, 412,  99

\reference{}{Romanowsky}, A.~J., {Douglas}, N.~G., {Arnaboldi}, M., {Kuijken}, K.,  {Merrifield}, M.~R., {Napolitano}, N.~R., {Capaccioli}, M., \& {Freeman},  K.~C. 2003, Science, 301, 1696

\reference{}{Rudnick}, G., {Labb{\'e}}, I., {F{\"o}rster Schreiber}, N.~M., {Wuyts}, S.,  {Franx}, M., {Finlator}, K., {Kriek}, M., {Moorwood}, A., {et al.} 2006, \apj, 650, 624

\reference{}{Rusin}, D. \& {Kochanek}, C.~S. 2005, \apj, 623, 666

\reference{}{Salpeter}, E.~E. 1955, \apj, 121, 161

\reference{}{Sandage}, A.~R. 1953, \aj, 58, 61

\reference{}{Scalo}, J.~M. 1986, Fundamentals of Cosmic Physics, 11, 1

\reference{}{Schiavon}, R.~P., {Caldwell}, N., \& {Rose}, J.~A. 2004, \aj, 127, 1513

\reference{}{Schlegel}, D.~J., {Finkbeiner}, D.~P., \& {Davis}, M. 1998, \apj, 500, 525

\reference{}{Schmidt}, M. 1959, \apj, 129, 243

\reference{}{Schweizer}, F. \& {Seitzer}, P. 1992, \aj, 104, 1039

\reference{}{Sirianni}, M., {Jee}, M.~J., {Ben{\'{\i}}tez}, N., {Blakeslee}, J.~P.,  {Martel}, A.~R., {Meurer}, G., {Clampin}, M., {De Marchi}, G., {et al.} 2005, \pasp, 117, 1049

\reference{}{Stanford}, S.~A., {Eisenhardt}, P.~R., \& {Dickinson}, M. 1998, \apj, 492, 461

\reference{}{Stanford}, S.~A., {Eisenhardt}, P.~R.~M., \& {Dickinson}, M. 1995, \apj, 450,  512

\reference{}{Stanway}, E.~R., {McMahon}, R.~G., \& {Bunker}, A.~J. 2005, \mnras, 359, 1184

\reference{}{Steidel}, C.~C., {Adelberger}, K.~L., {Shapley}, A.~E., {Erb}, D.~K., {Reddy},  N.~A., \& {Pettini}, M. 2005, \apj, 626, 44

\reference{}{Steidel}, C.~C., {Giavalisco}, M., {Pettini}, M., {Dickinson}, M., \&  {Adelberger}, K.~L. 1996, \apjl, 462, L17

\reference{}{Stolte}, A., {Brandner}, W., {Grebel}, E.~K., {Lenzen}, R., \& {Lagrange},  A.-M. 2005, \apjl, 628, L113

\reference{}{Stryker}, L.~L. 1993, \pasp, 105, 1081

\reference{}{Tafalla}, M., {Myers}, P.~C., {Caselli}, P., \& {Walmsley}, C.~M. 2004, \aap,  416, 191

\reference{}{Temi}, P., {Brighenti}, F., \& {Mathews}, W.~G. 2007, \apj, 660, 1215

\reference{}{Terlevich}, A.~I., {Caldwell}, N., \& {Bower}, R.~G. 2001, \mnras, 326, 1547

\reference{}{Thomas}, D., {Maraston}, C., {Bender}, R., \& {Mendes de Oliveira}, C. 2005,  \apj, 621, 673

\reference{}{Tinsley}, B.~M. 1980, Fundamentals of Cosmic Physics, 5, 287

\reference{}{Trager}, S.~C., {Faber}, S.~M., {Worthey}, G., \& {Gonz{\' a}lez}, J.~J.  2000a, \aj, 120, 165

\reference{}{Trager}, S.~C., {Faber}, S.~M., {Worthey}, G., \& {Gonz{\'a}lez}, J.~J.  2000b, \aj, 119, 1645

\reference{}{Tran}, K.-V.~H., {Franx}, M., {Illingworth}, G.~D., {van Dokkum}, P.,  {Kelson}, D.~D., {Blakeslee}, J.~P., \& {Postman}, M. 2007, \apj, 661, 750

\reference{}{Treu}, T. \& {Koopmans}, L.~V.~E. 2004, \apj, 611, 739

\reference{}{Treu}, T., {Stiavelli}, M., {Bertin}, G., {Casertano}, S., \& {M{\o}ller}, P.  2001, \mnras, 326, 237

\reference{}{Tumlinson}, J. 2007, \apj, 664, L63


\reference{}{van der Marel}, R.~P. 1991, \mnras, 253, 710

\reference{}{van der Marel}, R.~P. \& {van Dokkum}, P.~G. 2007,
\apj, 668, 756

\reference{}{van der Wel}, A., {Franx}, M., {van Dokkum}, P.~G., {Huang}, J., {Rix}, H.-W.,  \& {Illingworth}, G.~D. 2006a, \apjl, 636, L21

\reference{}{van der Wel}, A., {Franx}, M., {van Dokkum}, P.~G., {Rix}, H.-W.,  {Illingworth}, G.~D., \& {Rosati}, P. 2005, \apj, 631, 145

\reference{}{van der Wel}, A., {Franx}, M., {Wuyts}, S., {van Dokkum}, P.~G., {Huang}, J.,  {Rix}, H.-W., \& {Illingworth}, G.~D. 2006b, \apj, 652, 97

\reference{}{van Dokkum}, P.~G. 2005, \aj, 130, 2647

\reference{}{van Dokkum}, P.~G. \& {Franx}, M. 1996, \mnras, 281, 985

\reference{}---. 2001, \apj, 553, 90

\reference{}{van Dokkum}, P.~G., {Franx}, M., {Fabricant}, D., {Illingworth}, G.~D., \&  {Kelson}, D.~D. 2000, \apj, 541, 95

\reference{}{van Dokkum}, P.~G., {Franx}, M., {Kelson}, D.~D., \& {Illingworth}, G.~D.  1998a, \apjl, 504, L17

\reference{}{van Dokkum}, P.~G., {Franx}, M., {Kelson}, D.~D., {Illingworth}, G.~D.,  {Fisher}, D., \& {Fabricant}, D. 1998b, \apj, 500, 714

\reference{}{van Dokkum}, P.~G. \& {van der Marel}, R.~P. 2007, \apj, 655, 30

\reference{} Wilkins, S.~M., Trentham, N., \& Hopkins, A.~M. 2007,
\mnras, submitted

\reference{}{Worthey}, G. 1994, \apjs, 95, 107

\reference{}{Worthey}, G., {Faber}, S.~M., \& {Gonzalez}, J.~J. 1992, \apj, 398, 69

\reference{}{Xin}, Y. \& {Deng}, L. 2005, \apj, 619, 824

\reference{}{Yi}, S.~K., {Kim}, Y.-C., \& {Demarque}, P. 2003, \apjs, 144, 259

\reference{}{Yi}, S.~K., {Yoon}, S.-J., {Kaviraj}, S., {Deharveng}, J.-M., {Rich}, R.~M.,  {Salim}, S., {Boselli}, A., et al.\ 2005, \apjl, 619, L111

\reference{}{Zinn}, R.~J., {Newell}, E.~B., \& {Gibson}, J.~B. 1972, \aap, 18, 390

\reference{}Zoccali, M., Cassisi, S., Frogel, J.~A., Gould, A.,
 Ortolani, S., Renzini, A., Rich, R.~M., \& Stephens, A.~W.\ 2000,
 \apj, 530, 418

\end{references}
\end{document}